\documentclass[preprintnumbers,aps,prd,longbibliography,nofootinbib,nobibnotes,showpacs,amsmath,amssymb]{revtex4-1}
\usepackage{amsfonts}
\usepackage{mathrsfs}
\usepackage{epsfig}
\usepackage{graphicx}%
\usepackage{dcolumn}
\usepackage{amsmath}
\usepackage{color}
\usepackage{subfigure}

\begin{document}
\renewcommand{\baselinestretch}{1.3}
\newcommand\beq{\begin{equation}}
\newcommand\eeq{\end{equation}}
\newcommand\beqn{\begin{eqnarray}}
\newcommand\eeqn{\end{eqnarray}}
\newcommand\nn{\nonumber}
\newcommand\fc{\frac}
\newcommand\lt{\left}
\newcommand\rt{\right}
\newcommand\pt{\partial}

\title{Slowly rotating  black holes in the novel Einstein-Maxwell-scalar theory}

\author{Jianhui Qiu\footnote{jhqiu@nao.cas.cn}}
\affiliation{ National Astronomical Observatories, Chinese Academy of Sciences, Beijing {100101}, China}

\affiliation{School of Astronomy and Space Sciences, University of Chinese Academy of Sciences,
No. 19A, Yuquan Road, Beijing 100049, China}

\begin{abstract}
We investigate a slowly rotating black hole solution in a novel Einstein-Maxwell-scalar theory, which is prompted by the classification of general Einstein-Maxwell-scalar theories.  The gyromagnetic ratio of this black hole is calculated, and it increases as the second free parameter $\beta$ increases, but decreases with the increasing parameter $\gamma\equiv \frac{2 \alpha^{2}}{1+\alpha^2}$. In the Einstein-Maxwell-dilaton (EMD) theory, the parameter $\beta$ vanishes, but the free parameter $\alpha$ governing the strength of the coupling between the dilaton and the Maxwell field remains. The gyromagnetic ratio is always less than $2$, the well-known value for a Kerr-Newman (KN) black hole as well as for a Dirac electron. Scalar hairs reduce the magnetic dipole moment in dilaton theory, resulting in a drop in the gyromagnetic ratio. However, we find that the  gyromagnetic ratio of two can  be realized in this Einstein-Maxwell-scalar theory by increasing $\beta$ and the charge-to-mass ratio $Q/M$ simultaneously (recall  that the  gyromagnetic ratio of KN black holes is independent of $Q/M$).
The same situation also applies to the angular velocity of a locally non-rotating observer.  Moreover, we analyze the period correction for circular orbits in terms of charge-to-mass ratio, as well as the  correction of the radius of the innermost stable circular orbits.   It is found the correction increases with  $\beta$ but decreases with $Q/M$. Finally, the total radiative efficiency is investigated, and it can vanish once the effect of rotation is considered.

\end{abstract}









\maketitle



\section{Introduction}
Despite the great success of Einstein's general gravity (GR) in continued consistency with observations, there are compelling reasons to investigate theories of gravity beyond GR, ranging from the attempts  at quantum theories of gravity \cite{rovelli1998strings}, to the explanation of phenomena such as inflation \cite{cheung2008effective,weinberg2008effective}, dark matter \cite{famaey2012modified} and dark energy \cite{clifton2012modified,joyce2016dark}.

The Brans-Dicke gravity theory is one of the earliest suggested theories in the literature. It is expressed in the Jordan frame, where the scalar-tensor theory was initially developed according t o \cite{brans1961mach}. Brans-Dicke's theory describes  gravity by the metric tensor and a scalar field non-minimally coupled to gravity,  which O'Hanlon \cite{o1972intermediate}, Acharia and Hogan \cite{acharya1973equivalence} identified as the dilaton field. Based on the concept from particle physics, the dilaton field which manifests itself as the Nambu-Goldstone boson with broken scale -invariance, may mediate a limited range gravity \cite{fujii1971dilaton}. Indeed, by using the conformal transformations, one may extract the action of the Einstein-dilaton gravity theory from  scalar-tensor theory in Jordan's frame.  For a thorough understanding of scalar-tensor gravity and its relation to the dilaton field, we refer the readers to the monograph by Fujii and Maeda \cite{fujii2003scalar}.

 In the low-energy limit of several underlying quantum theories of gravity, the dilaton field, in conjunction with the axion field, which is deduced from string theory's low-energy limit, has produced intriguing results in inflationary cosmology and, more recently, in the acceleration of the universe \cite{sonner2006recurrent}. Exact static dilaton black hole solutions of Einstein-Maxwell-dilaton (EMD) gravity have been constructed by many authors \cite{gibbons1988black,garfinkle1991charged,cai1996black,cai1998topological}. However, exact rotating dilaton black hole solutions have been obtained only for some limited values of the dilaton coupling constant \cite{kunduri2005electrically,kunz2006rotating,brihaye2007black,sen1992rotating}, the most notable of which is the charged rotating Kerr-Sen solution, which takes into account both the dilaton and axion fields. In turn, the Kerr-Sen solution may be used to conduct an indirect test of string theory. Although the Kerr-Sen metric bears a strong resemblance to the Kerr-Newman metric, the inherent geometry of the two black holes varies significantly. The distinguished properties of the two spacetimes have been extensively studied in \cite{hioki2008hidden,pradhan2016thermodynamic,uniyal2017null,delgado2016violations}.
For general dilaton coupling constants, the characteristics of charged rotating dilaton black holes have been explored exclusively in the situation of infinitesimal tiny charge  \cite{casadio1997new} or angular momentum   \cite{horne1992rotating,shiraishi1992spinning,sheykhi2007thermodynamics}.
The slowly rotating black hole solutions in Einstein-Maxwell-Scalar theories in asymptotically flat or AdS spacetime have been widely researched (for references, see \cite{sheykhi2008asymptotically,ghosh2007slowly,ayzenberg2014slowly}).

Recently, the EMS models have attracted attentions once again owing to the studies on black hole spontaneous scalarization \cite{herdeiro2018spontaneous,fernandes2019spontaneous},
which arises from the earlier established spontaneous scalarization of neutron stars in the setting of scalar tensor theories. This kind of EMS model admits both a Reissner-Nordstr $\ddot{\textrm{o}}$m (RN) solution and a scalar solution. For a sufficiently large charge-to-mass ratio, however, the RN BH becomes unstable to scalar perturbations and dynamically evolves with a scalar field profile, making scalarization energetically advantageous.
Astefanesei et al. \cite{astefanesei2019einstein} proposed a classification of the BH solutions in EMS models, based on the behavior of the coupling function.
They considered the action
\begin{equation}
\mathcal{S}=\frac{1}{16 \pi} \int d^{4} x \sqrt{-g}\left(R-2 \partial_{\mu} \phi \partial^{\mu} \phi-K(\phi) F_{\mu \nu} F^{\mu \nu}\right)\;.
\end{equation}
EMS models are categorized into two categories based on whether or not the field equations admit the RN BH solution ($\phi(r)=0$, alternatively, $\left.K_{, \phi}(0) \equiv \frac{d K(\phi)}{d \phi}\right|_{\phi=0} = 0$). When RN BHs solve the field equations, the type is termed the scalarised-type; otherwise, the dilatonic-type.
For example, the case of $K(\phi)=e^{2\phi}$, i.e. the EMD theory, where RN BHs don't solve the field equations, presents a specific instance of the dilatonic-type.

Scalarised-type is further divided into two subclasses according to whether or not the scalar field profile is continuously connected with RN black holes (scalarised-connected-type or scalarised-disconnected type) by examining the linearization of the field equations for small $\phi$,
\begin{equation}
\left(\square-\mu_{\mathrm{eff}}^{2}\right) \phi=0, \quad \text { where } \mu_{\mathrm{eff}}^{2}=\left.\frac{F_{\mu \nu} F^{\mu \nu}}{4} \frac{d^{2} f(\phi)}{d \phi^{2}}\right|_{\phi=0} \,.
\end{equation}
If the condition of $\mu_{\mathrm{eff}}^{2}<0$ holds, then the scalarized BHs bifurcate from the RN BHs due to tachyonic instability, a process known as spontaneous scalarization. The scalarized BH reduces to the RN BH when $\phi$ satisfies  $\phi=0$, and then this type is termed as the scalarised-connected-type. A particular type, for instance, is $K(\phi)=e^{2\alpha\phi^2}$.

After examining the broad categories of EMS models, namely the dilatonic and scalarised type, one finds that dilaton black holes (allowed in dilatonic models) and RN black holes (allowed in scalarised models) both play critical roles. Then what will happen if the two mutually incompatible solutions are combined? To this end, we list the line elements of these two theories in the following. The line element of a dilaton black hole for $K(\phi)=e^{2\phi}$ is
\begin{equation}
ds^2=-\left(1-\frac{2M}{r}\right)dt^2+\frac{1}{\left(1-\frac{2M}{r}\right)}dr^2+r^2\left(1-\frac{Q^2}{Mr}\right)d\Omega \;,
\end{equation}
 while the line element of a RN BH, i.e. $K(\phi)=1$, is
\begin{equation}
ds^2=-	\left( 1-\frac{2M}{r}+\frac{Q^2}{r^2}\right)dt^2+\frac{1}{1-\frac{2M}{r}+\frac{Q^2}{r^2}}dr^2+r^2 d\Omega \,.
\end{equation}
A natural combination of the two line elements is \cite{yu2021constructing}
\begin{equation}
 \label{lineelement1}
ds^2=-	\left( 1-\frac{2M}{r}+\frac{\beta Q^2}{r^2\left(1-\frac{Q^2}{Mr}\right)}\right)dt^2+\frac{1}{1-\frac{2M}{r}+\frac{\beta Q^2}{r^2\left(1-\frac{Q^2}{Mr}\right)}}dr^2+r^2\left(1-\frac{Q^2}{Mr}\right) d\Omega \;,
\end{equation} 
where the angular component stems from  that of dilaton BHs, while the non-angular component combines those of dilaton BHs and RN BHs. $\beta$ is the second dimensionless free parameter that determines the coupling of the dilaton and Maxwell field. The corresponding action is then
\begin{equation}
\mathcal{S}_{1}=\frac{1}{16\pi}\int d x \sqrt{-g}\left(R-2 \nabla_{u} \phi \nabla^{\mu} \phi-\frac{2 e^{2 \phi}}{\beta+2+\beta e^{4 \phi}}F^2\right)\;,
\end{equation}
where $F_{\mu\nu}$ denotes the component of the Maxwell 2-form, $\phi$ a scalar field and $R$ the Ricci scalar.
Turimov et al. \cite{turimov2020test} have investigated the geodesic of this theory. We extended it by combining RN BHs and general dilaton BHs \cite{yu2021constructing} and by considering higher dimensions \cite{qiu2020constructing}. The present article is devoted to the study of the slowly rotating black hole in four-dimensional and asymptotically flat spacetime.  In this approach, we aim to get a better understanding of this theory. The article is organized in the following way.

In Section II, we briefly review this theory and derive the equations of motion for the slowly rotating black holes. Section III presents the numerical solution to black holes. In Section IV, we make a research on the properties of the slowly rotating black holes in two subsections, \textbf{A} and \textbf{B}. In subsection \textbf{A}, we study the angular momentum, the angular velocity of the event horizon, and the gyromagnetic ratio of the black holes. In subsection \textbf{B}, we compute the correction to geodesics due to the rotating effect. Finally, in Section V, we make a discussion and summarize our results.

 \section{ACTION AND THE EQUATIONS OF MOTION}

\begin{figure}[h]
 	\begin{center}
 		\includegraphics[width=8cm,height=6cm]{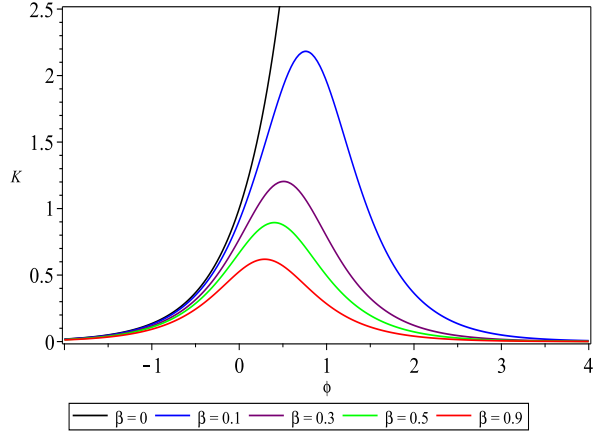}
 		\caption{The evolution of the coupling function $K(\phi)$. $K(\phi)$ is regular in the entire field space. The plots correspond to $\beta=0,\ \ 0.1,\  \ 0.3,\ \ 0.5,\ \ 0.9$, respectively, from top to bottom.}\label{figa}
 	\end{center}
 \end{figure}

The scalar field $\phi$ and the vector potential $A_{\mu}$ from action $\mathcal{S}_{1}$ are \cite{yu2021constructing}
 \begin{equation}
 	\phi=-\frac{1}{2}\ln\left(1-\frac{Q^2}{Mr}\right),
 	\label{phi1}
 \end{equation}
 and
 \begin{equation}
 	A_{\mu}=\left(A_t,A_r,A_{\theta},A_{\varphi}\right)=\left(\frac{Q}{r}+\frac{\beta Q}{2r}+\frac{\beta MQ}{2(Mr-Q^2)},0,0,0\right),
 \end{equation}
 respectively. To understand the role of coupling constant $\beta$, we display the evolution of  $K(\phi)$ against
 $\phi$ for various $\beta$ in Fig.~\ref{figa} \cite{yu2021constructing}.
 It is shown that with the increase of $\beta$, the effect of the Maxwell invariant diminishes and gravity takes over the electromagnetic interaction, allowing the electromagnetic field to be safely ignored. Apart from the situation of $\beta=0$ (i.e. EMD theory), the potential $K(\phi)$ exhibits an extreme at $\phi_0=\frac{1}{4}\ln  \left( {\frac {2+\beta}{\beta}} \right) $. Thus, if we make a transformation of $\phi \to \phi+\frac{1}{2}\ln(\phi_{0})$ on the coupling function $K(\phi)$ such that $K(\phi)\propto \frac{1}{cosh(2\phi)}$, we get a maximum for $K(\phi)$ at $\phi=0$. At the maximum, we have $dK/d\phi|_{\phi=0}=0$, implying that RN spacetime is the solution to action $\mathcal{S}_{1}$. However, as can be shown, solution (\ref{lineelement1}) is not identical to the RN solution, and the scalar field (\ref{phi1}) isn't trivial. {We'll show they constitute the second set of black hole solutions{\textemdash}the scalarised counterparts of the RN black holes according to the two Bekenstein-type identities in \cite{astefanesei2019einstein}.

 {The first identity is given by
 	\begin{equation}
 	\int\sqrt{-g}d^4x\left(K_{\phi\phi}\nabla_{\mu}\phi\nabla^{\mu}\phi+\frac{K_{\phi}^2}{4}F^2\right)=0\;.
 	\end{equation}
 	For a purely electric field, one has $F^2<0$, which suggests
 	\begin{equation}
K_{,\phi\phi}>0
 	\end{equation}	
should be satisfied in some region of $r$ outside the event horizon. Otherwise, the two terms of the integrand would always have the same sign, implying that the identity holds if and only if $\phi=0$.}

{The second identity is given by
 	\begin{equation}
 	\int\sqrt{-g}d^4x\left(\nabla_{\mu}\phi\nabla^{\mu}\phi+\frac{{\phi K_{,\phi}}}{4}F^2\right)=0\;.
 	\end{equation}
 	This reveals that for a pure electric field, the potential should satisfy the condition
 	\begin{equation}
 	\phi K_{,\phi}>0\;
 	\end{equation}
 	in some region of $r$ outside the event horizon.
 \begin{figure}[h]
 	\begin{center}
 		\includegraphics[width=8cm,height=6cm]{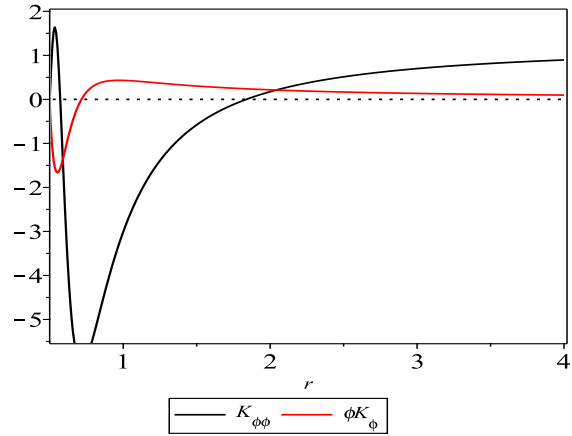}
 		\caption{The plots of $K_{,\phi\phi}$ and $\phi K_{,\phi}$  with respect to $r$, respectively. In some region outside the event horizon, one  has $K_{,\phi\phi}>0$ and $\phi K_{,\phi}>0$. Therefore, the theory has a scalarised black hole solution with a pure electric field.}\label{figb}
 	\end{center}
 \end{figure}
The graphs of $K_{,\phi\phi}$ and $\phi K_{,\phi}$ with $K=\frac{2 e^{2 \phi}}{\beta+2+\beta e^{4 \phi}}$ and $Q^2=0.5, M=1, \beta=0.2$ are shown in Fig.~\ref{figb}. It is apparent that in some range outside the event horizon (which is denoted by the zeros of metric component $g^{11}$), we always have $K_{,\phi\phi}>0$ and $\phi K_{,\phi}>0$. Then, we conclude that the theory has the scalarised black hole solution.

It is straightforward to demonstrate that $\mu_{\mathrm{eff}}^{2}>0$ for the electrical case. Therefore, it does not suffer the tachyonic
instability problem. In other words, the RN solution is stable to the scalar perturbations. According to \cite{astefanesei2019einstein}, this model is classified into scalarised-disconnected-type. It seems that the asymptotic value of scalar field in (\ref{phi1}) is set to zero, rather than arbitrary,
making the black hole solutions unnecessarily specific. Indeed, if we rescale $\phi\to \phi+\frac{1}{2}\ln(\phi_{0})$ and $A_{\mu}\to\left(\beta^2+2\beta\right)^{\frac{1}{4}}A_{\mu}$, where $\phi_{0}^2$ satisfies $\phi_{0}^2=\frac{2+\beta}{\beta}$ to make the new $K(\phi)$ achieve extreme at $\phi=0$, which is different from  FIG. 1, then the coupling constant $\beta$ disappears and the action becomes

\begin{equation}
\mathcal{S}_{2}=\frac{1}{4\pi}\int d^4x \sqrt{-g} \left(\frac{R}{4}-\frac{1}{4\cosh{2\phi}}F_{\mu\nu}F^{\mu\nu}-\frac{1}{2}\partial_\mu\phi \partial^\mu \phi\right)\ .
\label{action2}
\end{equation}
The corresponding solution is
 \begin{equation}
\begin{split}
&ds^2=-	\left( 1-\frac{2M}{r}+\frac{\beta \tilde{Q}^2}{r^2\left(1-\frac{\tilde{Q}^2}{Mr}\right)}\right)dt^2+\frac{1}{1-\frac{2M}{r}+\frac{\beta \tilde{Q}^2}{r^2\left(1-\frac{\tilde{Q}^2}{Mr}\right)}}dr^2+r^2\left(1-\frac{\tilde{Q}^2}{Mr}\right) d\Omega\;,\\
&\phi=-\frac{1}{2}\ln\left(\phi_0\left(1-\frac{\tilde{Q}^2}{Mr}\right)\right)\;,\\
&A_0=\left(\beta^2+2\beta\right)^{-\frac{1}{4}}\left(\frac{\tilde{Q}}{r}+\frac{\beta \tilde{Q}}{2r}+\frac{\beta M\tilde{Q}}{2(Mr-\tilde{Q}^2)}\right)\;,\\
&   \frac{\beta+2}{\beta}=\phi_{0}^2\;,
\end{split}
 \end{equation}
 where the physical electric charge of the black hole is given by $ Q=\frac{1}{4 \pi} \oint K(\phi) ^{*} F d \Omega=\left(\beta^2+2\beta\right)^{\frac{1}{4}}\tilde{Q}.$ Then, the asymptotic value of $\phi$ is not vanishing any more, and has an effect on the metric through the relationship between $\beta$ and $\phi_{0}$.

The coupling between the scalar field and the Maxwell field in $\mathcal{S}_{2}$ is $\frac{1}{cosh 2\phi}$ whereas in the dilaton theory is $e^{2\phi}$. This connection can be explained by Herdeiro and Oliveira's \cite{herdeiro2020electromagnetic} broader notion of electromagnetic duality: by some transformations, two different models are related by a non-trivial duality map. Let's begin from the Einstein-Maxwell-scalar class of model, whose action is given by
\begin{equation}
\mathcal{S}_0=\frac{1}{4\pi}\int d^4x \sqrt{-g} \left(\frac{R}{4}-\frac{f(\phi)}{4}F_{\mu\nu}F^{\mu\nu}+\frac{g(\phi)}{4}F_{\mu\nu}\tilde{F}^{\mu\nu}-\frac{1}{2}\partial_\mu\phi \partial^\mu \phi\right)\;,
\label{dualaction}
\end{equation}
where $\tilde{F}_{\mu\nu}$ denotes the Hodge dual of Maxwell 2-form and $f(\phi)\;, g(\phi)$ denote two unspecified coupling functions. The solution is described by
\begin{equation}
[{\bf g}, {\bf A}, \phi; f(\phi),g(\phi)]\; .
\label{dualsol}
\end{equation}
One can then establish an electromagnetic duality transformation defined by an angle $\theta$. $\mathcal{D}_\theta$ maps any solution (\ref{dualsol}) of a certain EMS model $\mathcal{S}_0$ to a different solution of a different (dual) model, within the same family as follows
\begin{equation}
[{\bf g}, {\bf A}, \phi; f(\phi),g(\phi)]  \stackrel{\mathcal{D}_\theta}{\longrightarrow} [{\bf g}, {\bf A}', \phi; f_\theta(\phi),g_\theta(\phi)] \ .
\label{orbit}
\end{equation}
In the case of
\begin{equation}
f(\phi)=e^{2\phi} \ , \qquad g(\phi)=0 \ ,
\end{equation}
which is first discussed by Gibbon and Maeda in\cite{gibbons1988black} and later by Garfinkle, Horowitz and Strominger\cite{garfinkle1991charged},
after the value of $\theta=\pi/4$ is taken, the model along the duality orbit has $f_\theta=1/\cosh{2\phi}$, $g_\theta=\tanh{2\phi}$ and its action is
\begin{equation}
\mathcal{S}_{\frac{\pi}{4}}'=\frac{1}{4\pi}\int d^4x \sqrt{-g} \left(\frac{R}{4}-\frac{1}{4\cosh{2\phi}}F'_{\mu\nu}F'^{\mu\nu}+\frac{\tanh{2\phi}}{4}F'_{\mu\nu}\tilde{F}'^{\mu\nu}-\frac{1}{2}\partial_\mu\phi \partial^\mu \phi\right) \;.
\label{dualaction2}
\end{equation}
Then, using equations (\ref{action2}) and (\ref{dualaction2}), we find that $\mathcal{S}_{2}$ differs from $\mathcal{S}_{\frac{\pi}{4}}'$ only in the axion term.

The generalization of $\mathcal{S}_{1}$ is represented by the action
 \begin{equation}
 \begin{split}
  &\mathcal{S}_{3}=\int d^{4} x \sqrt{g}\left(R-2 \nabla_{\mu} \phi \nabla^{\mu} \phi-K(\phi)  F^{2}\right)\;,\\
  &K(\phi)= \frac{e^{-\frac{2 \phi}{\alpha}}\left(\alpha^{2}+1\right)}{\left(\alpha^{2}+\beta+1\right) e^{-\frac{2 \phi\left(\alpha^{2}+1\right)}{\alpha}}+\beta \alpha^{2}}\;,
 \label{action3}
 \end{split}
 \end{equation}
after considering the general Einstein-Maxwell-dilaton solution \cite{horne1992rotating}.
The corresponding static solutions  are described by the line element $ds^2=-Udt^2+\frac{1}{U}dr^2+f^2d\Omega$, where
\begin{equation}\label{eq:u2a}
U=\left(1-\frac{b}{r}\right)\left(1-\frac{a}{r}\right)^{1-\gamma}+\frac{\beta Q^2}{f^2}\;,
\end{equation}
\begin{equation}\label{eq:u2b}
f=r\left(1-\frac{a}{r}\right)^{\frac{\gamma}{2}}\;,
\end{equation}
 and two other fields
 \begin{equation}
 \begin{split}
& \phi=-\frac{\alpha}{1+\alpha^2}\ln\left(1-\frac{a}{r}\right)\;,\ \ \ \ \  \\
&A_t=\frac{Q}{r}+\frac{1}{1+\alpha^2}\frac{\beta Q}{r}+\frac{\alpha^2}{1+\alpha^2}\frac{\beta Q}{r-a}\;.
 \end{split}
 \label{elec2}
 \end{equation}
It should be emphasized that $\gamma$ is introduced in order to simplify the notation and it is defined by
 \begin{eqnarray}
 \gamma \equiv \frac{2 \alpha^{2}}{1+\alpha^2}\;.
 \end{eqnarray}
 Here $a$ and $b$ are related to the mass $M$ and electric charge $Q$ of the black hole by
 \begin{equation}
 \label{eqmass}
 M=\frac{1}{2}\left[b+(1- \gamma) a\right]\;,\ \ \ \ \ \ Q^2=\left(1-\frac{\gamma}{2}\right)ab\;.
 \end{equation}
Observing the expressions of $U$ and $f$ in Eq.~(\ref{eq:u2a}) and Eq.~(\ref{eq:u2b}), we see the metric combines the dilaton part $\left(1-\frac{b}{r}\right)\left(1-\frac{a}{r}\right)^{1-\gamma}$ and RN part $\frac{\beta Q^2}{f^2}$ together.

 When $\beta=0$, the action $\mathcal{S}_{3}$ reduces to the model of gravity coupled to a
 Maxwell field and a dilaton field by Horne and Horowitz \cite{horne1992rotating}. When $\alpha=1$, the action $\mathcal{S}_{3}$ reduces to $\mathcal{S}_{1}$. One can again use a similar transformation like that from action $ \mathcal{S}_{1}$ to action $\mathcal{S}_{2}$ and  deduce the following action from $\mathcal{S}_{3}$
 \begin{equation}
 \mathcal{S}_{4}=\frac{1}{16\pi}\int d^{4} x \sqrt{g}\left(R-2 \nabla_{\mu} \phi \nabla^{\mu} \phi- \frac{2}{e^{-2\alpha\phi}+e^{\frac{2\phi}{\alpha}}}F^{2}\right)\;.
 \label{action4}
 \end{equation}

It's obvious that the analysis above on the case of  $\alpha=1$ can be extended to any $\alpha$ (see the analysis concerning $\mathcal{S}_1$ and $\mathcal{S}_2$).
It should be stressed once again, that in the solution of action  $\mathcal{S}_{2}$ and  $\mathcal{S}_{4}$, $\beta$ does not act as a free parameter but act as a third hair representing the scalar field's non-vanishing asymptotic value.

The above study fully illustrates the motivations for the novel Einstein Maxwell-scalar theory $\mathcal{S}_3$  \cite{yu2021constructing,qiu2020constructing,turimov2020test}, as well as the connection between it and the dilaton model. Now, let us extend our grasp of this theory by considering rotation, starting with the associated static solution of $\mathcal{S}_3$,  and concentrating on the search for a slowly rotating black hole solution in asymptotically flat spacetime. Varying the action $\mathcal{S}_3$ with respect to the metric, Maxwell and the scalar field, respectively, yields

\beq
\label{g0}
R_{\mu\nu}=2\nabla_\mu\phi\nabla_\nu\phi+2KF_{\mu\alpha}F_\nu^{\alpha}-\frac{K}{2}F^2g_{\mu\nu}\;,
\eeq

\beq
\label{F0}
\partial_\mu\left(\sqrt{-g}KF^{\mu\nu}\right)=0\;,
\eeq

\beq
\label{p0}
\nabla_\mu\nabla^\mu\phi-\frac{1}{4}\frac{\partial K}{\partial\phi}F^2=0\;.
\eeq

 We can solve equations (\ref{g0})-(\ref{p0}) to first order of the angular parameter $\epsilon$. We further assume, in accordance with \cite{horne1992rotating}, that the unique term in the metric that changes to the order of $O(\epsilon)$ is $g_{t\phi}$, that the scalar field does not change to the order of $O(\epsilon)$, and that $A_{\varphi}$ is the only component of the vector potential that changes to the order of $O(\epsilon)$. As a result, we suppose the metric has the  following form
\begin{eqnarray}
d{s}^2&=&-Ud{t}^2+\frac{1}{U}d{r}^2-2 \epsilon k(r) sin^2\theta dtd\varphi +
f^2d\Omega\;,
\end{eqnarray}
and the vector potential
\begin{equation}
A_{\mu}=(A_{t},0,0,-\epsilon QB(r)sin(\theta)^2)\;,
\end{equation}
with $A_{t}$ in equation(\ref{elec2}).
Inserting the metric, the Maxwell fields and the scalar field into the field equations leads to the perturbation equations
\begin{equation}
\label{R14}
\begin{aligned}
&\frac{-k(r)}{f(r)^{2}}+\frac{k(r) f^{\prime}(r) U^{\prime}(r)}{f(r)}+\frac{U(r) k^{\prime \prime}(r)}{2}\\=&-K(\phi)k(r) A_{t}^{\prime 2}-2 K(\phi)Q U(r) A_{t}^{\prime}(r) B^{\prime}(r)\;,
\end{aligned}
\end{equation}
and
\begin{equation}
\label{R14a}
\partial_{r}\left(-\left(k(r) A_{t}^{\prime}(r)+Q U(r) B^{\prime}(r)\right) K(\phi)\right)+\frac{2 Q B(r) K(\phi)}{f(r)^{2}}=0\;,
\end{equation}
where the prime denotes the derivative with respect to $r$.

Combing the equations of $R_{\theta\theta}$ and electromagnetic field
\begin{equation}
1-U f^{\prime 2}-U^{\prime} f f^{\prime}-U f f^{\prime \prime}=K A_{t}^{\prime 2} f^{2}\;,
\end{equation}

\begin{equation}
K(\phi)=-\frac{Q}{f^{2} A_{t'}^{\prime}}\;,
\end{equation}
we find equations (\ref{R14}) and (\ref{R14a}) can be reduced to
\begin{equation}
\label{eqk}
\frac{k^{\prime \prime} f^{2}}{2}=k f^{\prime 2}+k f f^{\prime \prime}+2 Q^{2} B^{\prime}\;,
\end{equation}
and
\begin{equation}
\label{eqB}
\frac{d\left(\frac{k }{f^{2}}\right)}{d r}-\frac{d\left(U B^{\prime} K(\phi)\right)}{d r}+\frac{2 B K(\phi)}{f^{2}}=0\;,
\end{equation}
respectively.
Integrate equation(\ref{eqk}), and then we achieve
\begin{equation}
\label{eqk2}
f^{2} k^{\prime}-\left(f^{2}\right)^{\prime} k=4 Q^{2} B+ { const }\;.
\end{equation}
Inserting it into equation (\ref{eqB}), we obtain
\begin{equation}
\label{eqB2}
\frac{4Q^2B+const}{f^4}-\frac{d\left(U B^{\prime} K(\phi)\right)}{d r}+\frac{2 B K(\phi)}{f^{2}}=0\;.
\end{equation}
We have now obtained two main equations (\ref{eqk2}) and(\ref{eqB2}). Before we solve the two equations, we must first consider the boundary conditions, as seen in the next section.

\section{Methodology for solving the equation}
Observing equations (\ref{eqk2}) and(\ref{eqB2}), one finds the value of the \emph{const} hasn't been determined, and we'll see later that the angular momentum ofspacetime is proportional to \emph{const}. As usual, we require the solutions have the behavior of $k(r) \to O(1/r) $ and $B(r) \to 1/r $ when $r\to \infty$.
Solving equation (\ref{eqk2}), we obtain
\begin{equation}
\label{ksolution}
k=f^2\int \frac{4 Q^{2} B+c o n s t}{f^{4}}+c_{1} f^{2}.
\end{equation}
Given that $k(r) \to O(1/r) $  at infinity, the  asymptotic behavior requires that $c_{1}$ must vanish.

Now we focus on the equation of $B(r)$, i.e. equation (\ref{eqB2}). First, we search for the Frobenius series solution of $B(r)$ at infinity. Using the reciprocal substitution $r\equiv1/y$, we transform equation (\ref{eqB2}) into
\begin{equation}
\label{eqB3}
\frac{4 Q^{2} B+\operatorname{const}}{f^{4}}+y^{2} \frac{\partial\left(-U y^{2} \frac{\partial B}{\partial y} K(\phi)\right)}{\partial y}+\frac{2 B K(\phi)}{f^{2}}=0\;.
\end{equation}
For the sake of calculation of the coefficients in the series, we multiply equation (\ref{eqB3}) by $f^4/K(\phi)^2$, and then the equation is transformed into
\begin{equation}
\label{eqB4}
\frac{4 Q^{2} B+\operatorname{const}}{K(\phi)^{2}}-\frac{y^{4} f^{4} U}{K(\phi)} \frac{d^2 B}{d y^{2}}-\frac{d\left(U y^{2} K(\phi)\right)}{d y} \frac{f^{4}}{K(\phi)^{2}} y^{2} \frac{d B}{d y}+\frac{2 B f^{2}}{K(\phi)}=0\;.
\end{equation}
By expanding the coefficients of $\frac{d^2B}{dy^2}$, $\frac{dB}{dy}$ and $B$ at $y=0$, one finds they are in the order of $O(1)$, $O(1/y)$ and $O(1/y^2)$, respectively.  This demonstrates $y=0$ is a regular singular point of the equation. Two roots of the indicial equation of the homogeneous equation are $-2$ and $1$, respectively. Therefore, we will omit the exponent of $-2$ since $B(y)=O(y)$ is required when $y \to 0$ .
As a result, we substitute
\begin{equation}
B(y)=y \sum_{n=0}^{\infty} c_{n} y^{n}\;,
\end{equation}
into equation (\ref{eqB4}) and the coefficient of $y^0$ term gives
\begin{equation}
{\beta}^{2}{\it const}+ \left(  \left(  \left( -\gamma+3 \right) a+3\,b
\right) {\it c_{0}}-4\,{\it c_{1}}+2\,{\it const} \right) \beta+ \left(
\left( -2\,\gamma+3 \right) a+3\,b \right) {\it c_{0}}-4\,{\it c_{1}}+{
	\it const}=0\;.
\end{equation}
As in customary, we need $c_0=1$(required by $B(r) \to 1/r $), but there are still two variables to be determined, namely ${const}$ and $c_1$. The equations of higher orders will introduce  $c_2$, $c_3$, and so on, indicating that the equations of coefficients are not closed.

When $\beta=0$ (i.e. EMD theory), Sheykhi et al. provided a particular solution, $B(r)=1/r$ in \cite{sheykhi2008asymptotically}.
However, since finding an exact solution to $B(r)$ for an arbitrary value of $\beta$  is difficult, we must resort to numerical approaches, where one should set the boundary condition first.
In \cite{barausse2016slowly}, Barausse et al. made a study on slowly rotating black holes in Einstein-aether theory where they impose the condition that the solutions are regular everywhere, except for their center (singularity of the black hole). Equation (\ref{eqB4}) exhibits apparent singularities on the horizon of the black holes as defined by $U=0$, i.e.
\begin{equation}
\label{quad-eq}
(r_h-a)(r_h-b)+\beta Q^2=0\;.
\end{equation}
If the solution is to be regular there, $B'(r)$ and $B(r)$ should satisfy the following equation on the horizon $y_h$ (or equivalently, $1/r_h$),
\begin{equation}
\label{bc1}
\left[\frac{4 Q^{2} B+\operatorname{const}}{K(\phi)^{2}}-\frac{d\left(U y^{2} K(\phi)\right)}{d y} \frac{f^{4}}{K(\phi)^{2}} y^{2} \frac{d B}{d y}+\frac{2 B f^{2}}{K(\phi)}\right]_{y=y_h}=0\;.
\end{equation}
Combining this condition and the asymptotic behavior of $B(r)$ at infinity, we can specify the exact value of \emph{ const}. However, it should be noticed that there is not one but two horizons in the general case from the quadratic equation (\ref{quad-eq}). Actually, there will be two horizons ($r_h > a $) as long as $0<\beta Q^2<\frac{(b-a)^2}{4}$. In general, when two horizons exist, one would  anticipate that two singularities would appear in the equation. As mentioned before, to preserve the regularity on the horizon, we must implement the local regularity condition. Thus, one has to impose two local conditions in the presence of two horizons. However, by applying one regular condition alone, the solution is already specified without any tuning of other parameters in order to impose further regularity conditions. Therefore, it is sufficient to impose the regular requirement on just the outermost horizon although the solutions with many horizons will display several singularities on the horizons. This is acceptable because the outermost horizon can be rendered regular by using the usual regularity condition while the remaining singularities are hidden in the outermost horizon. Therefore, the subscript $y_h$ in (\ref{bc1}) represents the reciprocal radius of the outermost horizon.

In order to solve the equations numerically, we first set $const=1$ by rescaling $B(r)$, and then impose the boundary condition (\ref{bc1}) and $B(0)=0$ where $y_h= \left( a/2+b/2+1/2\,\sqrt {2\,ab\beta\,\gamma-4\,ab\beta+{a}^{2}-2\, ab+{b}^{2}} \right)^{-1}$.
Once the numerical solution is obtained, we can extract $B'(0)$ from it. Then we get the exact value of  $const=1/B'(0)$ since we require $B(y)\sim y$ at $y=0$.

\section{Properties of the slowly rotating black holes}

In this section, we investigate the case of $\beta \geq 0$, where the sign of $K(\phi)$ is always positive in the whole spacetime. Otherwise, there would exist regions where $K(\phi)<0$, in which the Maxwell field would become a phantom field and won't be considered in this article.

\subsection{angular momentum, gyromagnetic ratio, and angular velocity}
Now, let us consider the influence of $\beta$ on $const$. We fix $a=1, b=3$ and set  $\beta=0, 0.2, 0.4, 0.6, 0.8$ each time, and then conduct research on the relationship between $const$ and $\gamma$, which is displayed in Fig.~\ref{fig1} (As for the physical meaning of $a$ and $b$, keep in mind that they are connected with the mass $M$ and electric charge $Q$ through equation (\ref{eqmass})).
\begin{figure}[htbp]
	\label{const-gamma.eps}
	\centering
	\includegraphics[width=8cm,height=6cm]{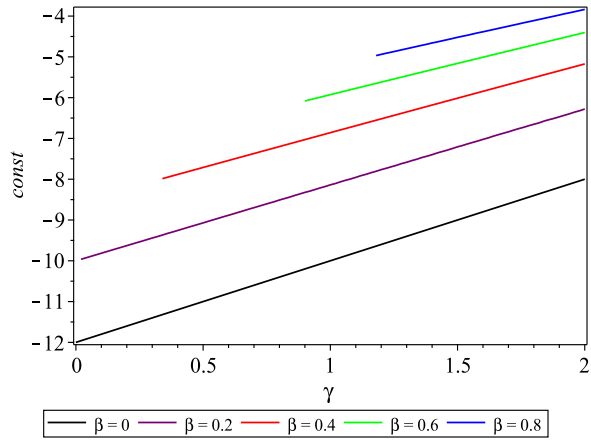}
	\caption{The relation between \emph{const} and $\gamma$. The parameters are $a=1$, $b=3$.}
	\label{fig1}
\end{figure}
The figure shows that several curves cannot be extended into $\gamma=0$ since we  require $0<\beta Q^2<\frac{(b-a)^2}{4}$; otherwise, there will be naked singularity which violates the cosmic censorship conjecture. Considering that $\gamma \equiv\frac{2 \alpha^2}{\alpha^2 +1}$, one will find the range of $\gamma$ is $[0,2)$. Once  the numerical value of $const$ is obtained, we may get a truncated series solution for $B(y)$. As a final check, our numerical solution is compared with the series solution to $B(y)$ in Fig.~\ref{fig2}. 

\begin{figure}[htbp]
	\centering
	\includegraphics[width=8cm,height=6cm]{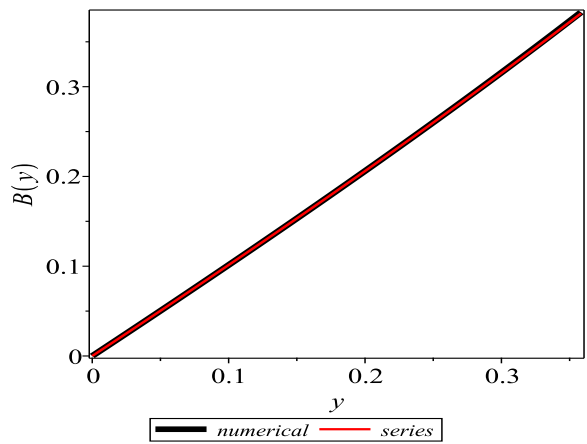}
	\caption{The numerical and the series solution for the relation $B$ and $y$. The parameters are $a=1$, $b=3$, $\gamma=1$, $\beta=1/4$. The truncated series solution is $y+0.12056 y^{2}+0.12404 y^{3}+0.12748 y^{4}$. }
	\label{fig2}
\end{figure}

Next, we calculate the angular momentum of the slowly rotating black hole. It can be calculated using the method provided by Brown and York \cite{brown1993quasilocal}. Conserved charges such as angular momentum are defined using the surface stress tensor and Killing vector fields on the boundary of spacetime. The boundary of $\Sigma$ is $\mathscr{B}$ with the metric $\sigma_{i j}$, and the product of $\mathscr{B}$ with segments of timelike world lines orthogonal to $\Sigma$ at $\mathscr{B}$ is denoted as $^3\mathscr{B}$. The surface stress energy tensor is then defined by
\begin{equation}
\tau^{a b}=\frac{1}{8 \pi}\left(\Theta^{a b}-\Theta \gamma^{a b}\right),
\end{equation}
which is derived from the variation of the action with respect to the metric $\gamma^{a b}$. We decompose the boundary metric $\gamma^{a b}$ into the standard ADM form,
\begin{equation}
\gamma_{a b} d x^{a} d x^{b}=-N^{2} d t^{2}+\sigma_{i j}\left(d \varphi^{i}+V^{i} d t\right)\left(d \varphi^{j}+V^{j} d t\right)\;,
\end{equation}
where we have chosen the two-surface as a two-sphere, and the coordinates $\varphi^i$ are the angular variables parameterizing the hypersurface of constant $r$ around the origin. $\sigma$ is the determinant of $\sigma_{i j}$. $N$ and $V^i$ are the lapse and shift functions, respectively.

Suppose $^3\mathscr{B}$ possesses an isometry concerned with Killing vector field $\xi$. then, the corresponding conserved charge is defined by
\begin{equation}
\int_{\mathcal{B}} d^{2} \varphi \sqrt{\sigma} \tau_{a b} n^{a} \xi^{b}\;,
\end{equation}
where $n^a$ is the normal vector of $\mathscr{B}$ and  is tangent to $^3\mathscr{B}$.
The conserved charge associated with the rotational Killing vector field $\frac{\partial}{\partial \varphi}$ is the angular momentum,

\begin{equation}
J=\int_{\mathcal{B}} d^{2} \varphi \sqrt{\sigma} \tau_{a b} n^{a}\left(\frac{\partial}{\partial \varphi}\right)^b\;.
\end{equation}
Expanding $\tau_{a b} n^{a}\left(\frac{\partial}{\partial \varphi}\right)^b$ and collecting the terms of order $O(\epsilon)$, we obtain

\begin{equation}
	\tau_{a b} n^{a}\left(\frac{\partial}{\partial \varphi}\right)^b=\sin^2\theta\frac{\left(-2kf'(r)+f(r)k'(r)\right)\epsilon}{16\pi f(r)}\;.
\end{equation}
Perform the integration and now the angular momentum equals to
\begin{equation}
J=\lim _{r \rightarrow \infty} -\frac{1}{6}\left(Q^{2} B+\operatorname{const}\right)\epsilon\;.
\end{equation}
Taking account of $B(r)=O(1/r)$ at infinity, the angular momentum equals to $-\epsilon \frac{const}{6}$.
Angular momentum can also be calculated using the Komar angular momentum and the same result obtained (for details see \cite{gourgoulhon20123+}). When $\beta=0$, we have $const=(2\gamma-3)a-3b$ and the angular momentum is
\begin{equation}
J=\frac{\epsilon}{2}\left(b+\frac{3-\alpha^{2}}{3\left(1+\alpha^{2}\right)} a\right)\;,
\end{equation}
which is the same as the value in \cite{sheykhi2008asymptotically}.

Once we get the numerical value of $const$, we can calculate the gyromagnetic ratio of the black hole. One of the remarkable facts about a Kerr-Newman black hole is that it has the same gyromagnetic ratio as an electron in the Dirac theory, $g=2$. Scalar fields, such as the dilaton field, modify the value of gyromagnetic ratio of the black hole, as a result, it doesn't possess the gyromagnetic ratio of $g=2$ in \cite{horne1992rotating}.
We will now examine the effect of $\beta$ on the gyromagnetic ratio $g$. The magnetic dipole moment for this asymptotically flat, slowly rotating black hole can be defined as
\begin{equation}
\mu=Q\epsilon=g\frac{QJ}{2M}\;.
\end{equation}
Substituting $M=\frac{1}{2} \left(b+(1-\gamma)a\right)$ (\ref{eqmass}) and $J=-\frac{\epsilon}{6}const $ into above equation, we
obtain $g=-6\frac{ \left(b+(1-\gamma)a\right)}{const}$.
\begin{figure}[htbp]
	\label{g1}
	\centering
	\includegraphics[width=8cm,height=6cm]{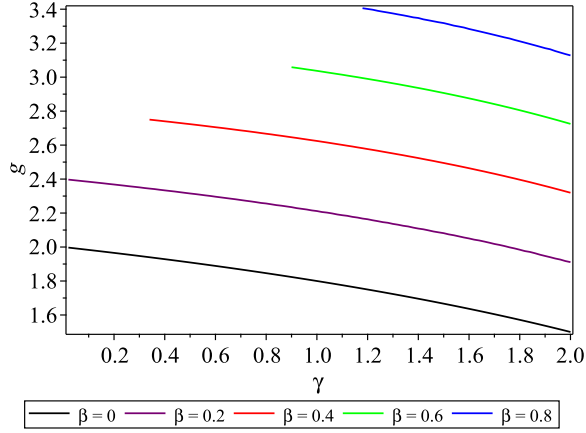}
	\caption{The gyromagnetic ratio with respect to $\gamma$. The parameters are $a=1$, $b=3$. }
	\label{fig3}
\end{figure}
In the case of $\beta=0$, i.e. the four dimensional  dilaton black hole, we have  $const=(2\gamma-3)a-3b$, and the gyromagnetic ratio is
\begin{equation}
\label{g2}
g=2-\frac{4 \alpha^{2} a}{\left(3b-a\right) \alpha^2+3a+ 3b}\;.
\end{equation}
 For a dilaton black hole, $r=a$ denotes a curvature singularity and  $r=b$ represents an event horizon. Thus, $a<b$ is ensured by cosmic censorship. When the denominator of the fraction of (\ref{g2}) is examined, it's found to be always positive, indicating that the gyromagnetic ratio has an upper limit of 2, the well-known gyromagnetic ratio of the Kerr-Newman black hole. By including scalar hair, the gyromagnetic ratio is suppressed. This has been shown in \cite{delgado2016kerr}, which extends the Kerr-Newman black hole to include scalar hair.

Turning to the behavior of the gyromagnetic ratio of the slowly rotating black hole corresponding  to $\mathcal{S}_3$, {we plot the gyromagnetic ratio $g$ versus $\gamma$ (recall  that $\gamma \equiv \frac{2 \alpha^{2}}{1+\alpha^2}$) in Fig.~\ref{fig3}. From this figure, we find  that the gyromagnetic ratio decreases with $\gamma$ or $\alpha$ when the value of $\beta$ is fixed. In the case of $\beta=0$ (black line) \cite{sheykhi2008asymptotically}, the gyromagnetic ratio decreases starting from $2$, the value for the Kerr-Newman black hole. However, as seen in the figure, when  $\beta    \neq0$, the gyromagnetic ratio can surpass 2. Taking $\gamma=0$ as an example, we have $B(r)=1/r$ and in this case $const=-\frac{(3a+3b)}{\beta+1}$, implying that gyromagnetic ratio is $2(\beta+1)$.  This is a specific example that the non-vanishing positive $\beta$ increases the gyromagnetic ratio. Based on the analysis above, the Kerr-Newman black hole's gyromagnetic ratio may be obtained in the Einstein-Maxwell-scalar black hole by simultaneously raising the values of $\beta$ and $\gamma$.

For the sake of simplicity, we shall restrict ourselves to the case of $\gamma=1$, i.e, $\alpha=1$. Then the corresponding action is $\mathcal{S}_{1}$. Expanding the expression of $K(\phi)=\frac{2 e^{2 \phi}}{\beta+2+\beta e^{4 \phi}}$ in terms of $\beta$
\begin{equation}
K(\phi)=\frac{2 e^{2 \phi}}{\beta+2+\beta e^{4 \phi}}=e^{2 \phi}-\frac{\left(e^{4 \phi}+1\right) e^{2 \phi} \beta}{2}+\frac{e^{2 \phi}\left(e^{4 \phi}+1\right)^{2} \beta^{2}}{4}+...\;,
\end{equation}
one finds that $K(\phi)$ is actually the combination of infinite distinct dilaton couplings. In Fig.~\ref{fig4}, we show the gyromagnetic ratio of the slowly rotating black hole in relation to $\frac{Q}{M}$ for various values of $\beta$ . The black line denotes the usual dilaton field $K(\phi)=e^{2\phi}$, whereas the other lines represent the situations where $\beta \neq 0 $. We find the upper limit for $\frac{Q}{M}$ with fixed $\beta$ is
\begin{equation}
\sqrt{(2 \beta+2)-2 \sqrt{\beta(\beta+2)}}\;,
\end{equation}
in order to ensure that the singularity $r=a$ of the black hole is not naked. It indicates that the gyromagnetic ratio increases with increasing $\beta$ but decreases with increasing $\frac{Q}{M}$, which means we can achieve the same ratio as Kerr-Newman black holes by simultaneously increasing $\beta$ and charge-to-mass ratio.

\begin{figure}[htbp]
	\centering
	\includegraphics[width=8cm,height=6cm]{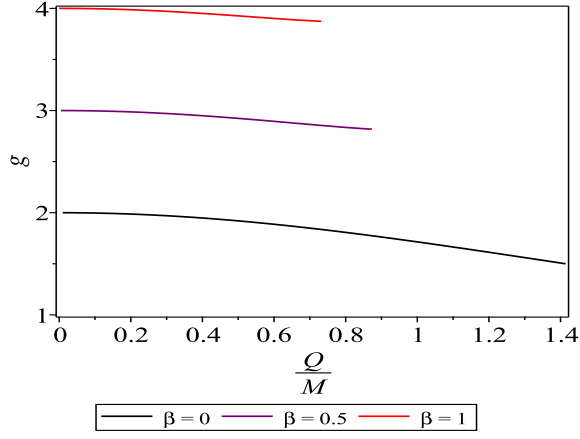}
	\caption{The gyromagnetic ratio with respect to $\frac{Q}{M}$ for various values of $\beta$.}
	\label{fig4}
\end{figure}

Next, we will investigate the angular velocity of the black hole horizon. The coordinate angular velocity of a locally
non-rotating observer is defined by
$\Omega=-\frac{g_{tr}}{g_{rr}}=\epsilon \frac{k(r)}{f(r)^2}$. It is known that one of the important quantities is the angular velocity on the horizon $\Omega_h=\Omega(r=r_h)$, which affects the region where the super-radiation occurs in the black hole background \cite{misner1972stability,zel1972amplification,starobinskii1973amplification}.

Because the angular velocity of the horizon is equal to $\epsilon \frac{1}{r_{h}^2}$ for a slowly rotating charged Kerr black hole, we can make the angular velocity dimensionless as follows $\tilde{\Omega}_h=\Omega_h/(\frac{\epsilon}{r_h^2})=\epsilon \frac{k(r_h)}{f(r_h)^2}/(\frac{\epsilon}{r_h^2})=\frac{k(r_h)r_h^2}{f(r_h)^2}$.  In Fig.~\ref{fig5}, we exhibit the dimensionless angular velocity $\tilde{\Omega}_h$ as the function of the charge-to-mass ratio.
\begin{figure}[htbp]
	\centering
	\includegraphics[width=8cm,height=6cm]{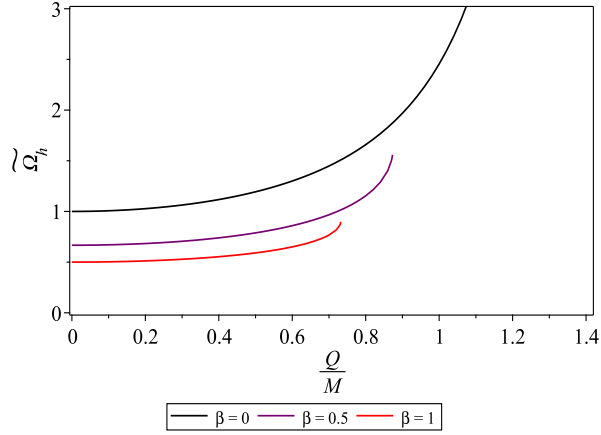}
	\caption{The dimensionless angular velocity on the horizon with respect to charge-to-mass ratio for different values of $\beta$. }
	\label{fig5}
\end{figure}
As can be seen, the dimensionless angular velocity on the horizon increases with increasing charge-to-mass ratio for fixed $\beta$. However, since the angular velocity decreases with $\beta$ when $\frac{Q}{M}$ is fixed, one may obtain the same angular velocity $\tilde{\Omega}_h$ as the Kerr-Newman black holes by simultaneously increasing the value of $\beta$ and $\frac{Q}{M}$.

\subsection{The innermost stable circular orbit, the radiation efficiency and their corrections}
In this section, we will focus on the circular orbits in the equatorial plane in order to investigate the geometry of the spacetime above.
In the stationary and axially symmetric spacetime, one can find the equations of motion for geodesics in the form \cite{shibata1998innermost}

\begin{equation}
\begin{array}{l}
\dot{t}=\frac{-E g_{\phi \phi}-L_{z} g_{t \phi}}{-g_{t \phi}^{2}+g_{t t} g_{\phi \phi}}\;, \\
\dot{\phi}=\frac{-E g_{t \phi}-L_{z} g_{t t}}{g_{t \phi}^{2}-g_{t t} g_{\phi \phi}}\;, \\
g_{r r} \dot{r}^{2}+g_{\theta \theta} \dot{\theta}^{2}=V_{e f f}\left(r, \theta ; E, L_{z}\right)\;,
\end{array}
\end{equation}
with the effective potential given by
\begin{equation}
V_{\rm e f f}(r)=\frac{E^{2} g_{\phi \phi}+2 E L_{z} g_{t \phi}+L_{z}^{2} g_{t t}}{g_{t \phi}^{2}-g_{t t} g_{\phi \phi}}-1\;,
\end{equation}
where the overhead dot stands for the derivative with respect to the affine parameter, and the constants $E$ and $L_z$ correspond to the conserved energy and the (z-component of) orbital angular momentum of the particle, respectively.

For simplicity, we put the orbits on the equatorial plane. With the constraint that $\theta=\frac{\pi}{2}$, one finds the effective potential $V_{\rm eff}(r)$ must satisfy
\begin{equation}
V_{\rm eff}(r)=0\;,\ \ \ \  \frac{dV_{\rm eff}(r)}{dr}=0\;,
\end{equation}
in order that the circular orbit in the equatorial plane is stable.

Solving the above equations, one obtains
\begin{equation}
\label{isco1}
\begin{array}{l}
E=\frac{-g_{t t}-g_{t \phi} X}{\sqrt{-g_{t t}-2 g_{t \phi} X-g_{\phi \phi} X^{2}}}\;, \\
L_{z}=\frac{g_{t \phi}+g_{\phi \phi} X}{\sqrt{-g_{t t}-2 g_{t \phi} X-g_{\phi \phi} X^{2}}}\;, \\
X=\frac{d \phi}{d t}=\frac{-g_{t \phi, r}+\sqrt{\left(g_{t \phi, r}\right)^{2}-g_{t t, r} g_{\phi \phi, r}}}{g_{\phi \phi, r}}\;.
\end{array}
\end{equation}
Then the corrections to energy, angular momentum, and period up to the first order of $\epsilon$ are given by
\begin{equation}
E^{2}=\frac{U^{2}\left(f^{2}\right)^{\prime}}{U \left(f^{2}\right)^{\prime}-f^{2} U^{\prime}}+\epsilon \cdot \frac{2 U \sqrt{U^{\prime} \left(f^{2}\right)^{\prime}} f^{2}\left(k^{\prime} U-k U^{\prime}\right)}{\left(-U^{\prime} f^{2}+U\left(f^{2}\right)^{\prime}\right)^{2}}\;,
\end{equation}

\begin{equation}
L^{2}=\frac{f^{4} U^{\prime}}{U\left(f^{2}\right)^{\prime}-U^{\prime} f^{2}}+\epsilon \cdot \frac{-2 f^{2} U \sqrt{U^{\prime}\left(f^{2}\right)^{\prime}}\left(-k^{\prime} f^{2}+\left(f^{2}\right)^{\prime} k\right)}{\left(-U^{\prime} f^{2}+U\left(f^{2}\right)^{\prime}\right)^{2}}\;,
\end{equation}
and
\begin{equation}
T^{2}=T_0^2+\epsilon T_1^2=\frac{4 \pi^{2}\left(f^{2}\right)^{\prime}}{U^{\prime}}-4 \pi^{2} \epsilon \cdot \frac{2\left(f^{2}\right)^{\prime} k^{\prime}}{\sqrt{U^{\prime}\left(f^{2}\right)^{\prime}} U^{\prime}}\;.
\end{equation}
We don't bother showing the detailed formula for energy, angular momentum, and the orbital period since they are rather lengthy.
The relative correction of the period is
\begin{equation}
\Delta T^{2}=\frac{T_{1}^{2}}{T_{0}^{2}}=-\frac{2 k^{\prime}}{\sqrt{U^{\prime}\left(f^{2}\right)^{\prime}}}\;,
\end{equation}
which is shown in Fig.~\ref{fig6}.
\begin{figure}[htbp]
	\centering
	\includegraphics[width=8cm,height=6cm]{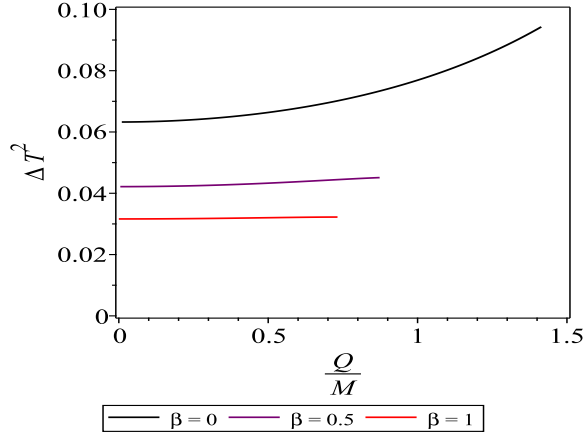}
	\caption{The corrected period with respect to $\frac{Q}{M}$ for different values of $\beta$. The radius of circular orbit is set to $10M$ and $\epsilon=1$.}
	\label{fig6}
\end{figure}
As can be shown, for fixed values of radius for circular orbits and fixed $\beta$, the relative correction of the period increases with increasing $\frac{Q}{M}$. The correction becomes smaller and smaller with the increase of $\beta$.

The innermost stable circular orbit (ISCO) of the particle around the black hole is given by the equation $V_{{\rm eff},rr}=0$, i.e.
\begin{equation}
\label{isco2}
{E^{2} g_{\phi \phi}^{\prime \prime}+2 E L_{z} g_{t \phi}^{\prime \prime}+L_{z}^{2} g_{t t}^{\prime \prime}}={(g_{t \phi}^{2})^{\prime \prime}-(g_{t t} g_{\phi \phi})^{\prime \prime}}\;.
\end{equation}
Substituting the expressions of $E$, $L_z$, $g_{tt}$, $g_{t\phi}$ and $g_{\phi \phi}$ into Eq.~(\ref{isco2}), we obtain the equation denoted by $P(R,\epsilon)=0$, where $R$ denotes the radius of ISCO of the rotating black hole. Assuming $R_{0}$ is the radius of ISCO in the corresponding static spacetime, i.e. $P(R_{0},0)=0$, the correction to the radius of ISCO up to the first order of $\epsilon$ is then $R_{0}+\epsilon R_{1}$. Here, we have $R_1=-\frac{P_2(R_{0},0)}{P_1(R_{0},0)}$.
$P_1(R_{0},0)$ and $P_2(R_{0},0)$ denote the derivatives of $F$ with respect to the first and second variables, respectively.

Again, we don't bother giving the expression for $R_{0}$ since it is the root of a quartic equation. We only display the radius of the innermost stable circular orbits concerning $\frac{Q}{M}$ for various $\beta$ in Fig.~\ref{fig7}(a). The relative correction, denoted by $\frac{R_1}{R_{0}}$ is shown in Fig.~\ref{fig7}(b).
 \begin{figure}[htbp]
	\centering
	\subfigure[]{
		\includegraphics[width=8cm,height=6cm]{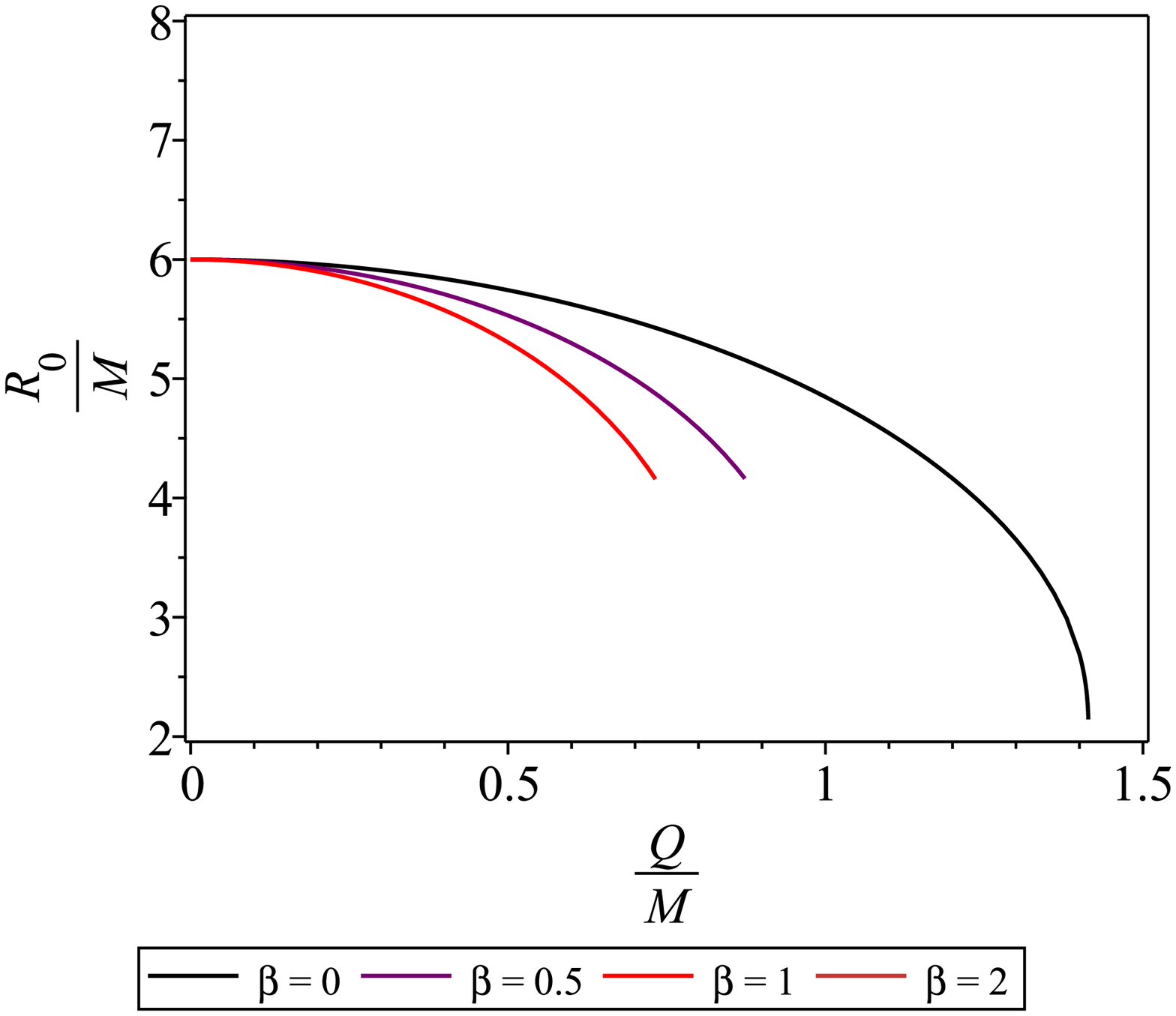}}
	\subfigure[]{
		\includegraphics[width=8cm,height=6cm]{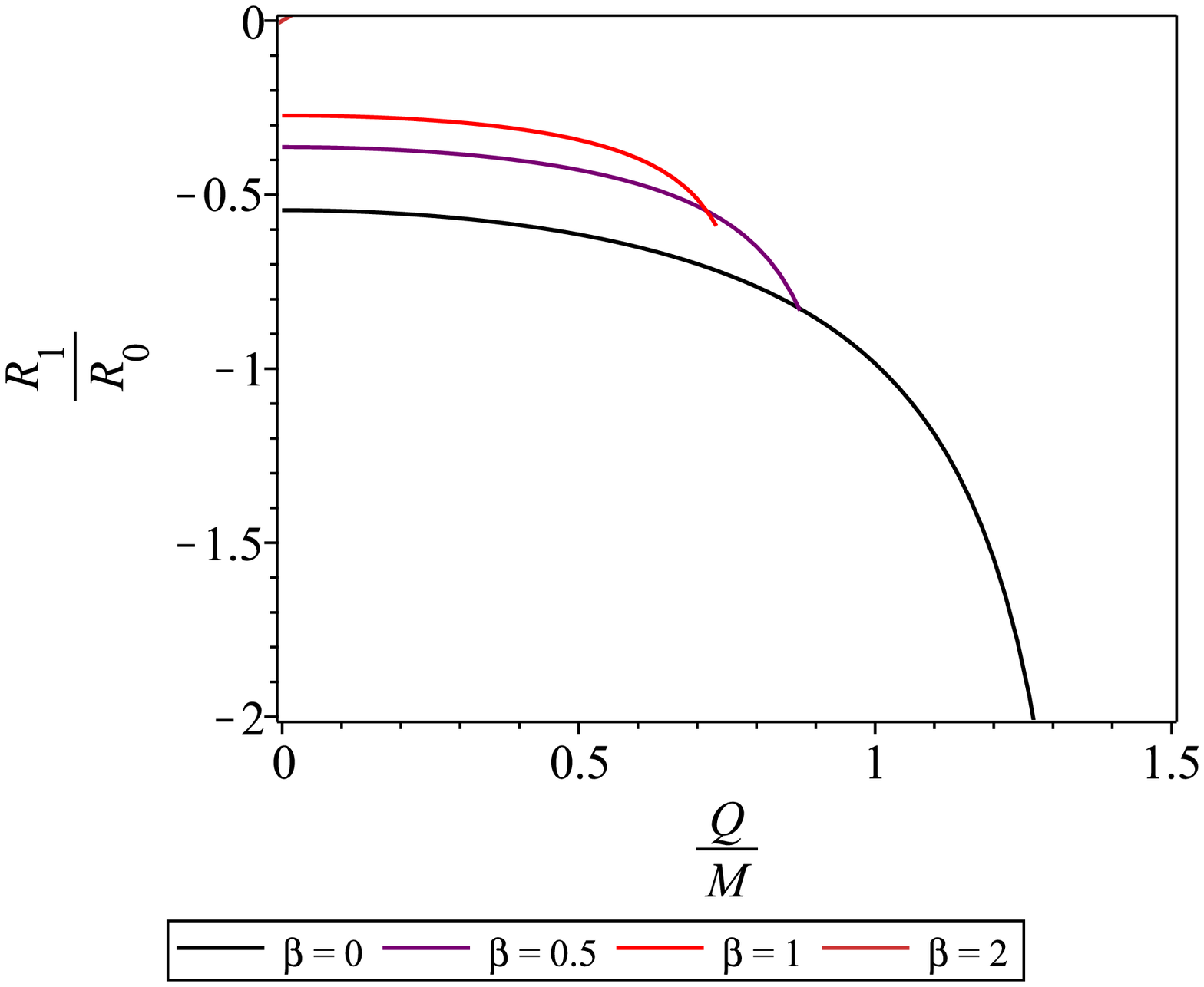}}\\
	\caption{(a) The dimensionless radius of the innermost stable circular orbit with respect to $\frac{Q}{M}$ for different values of $\beta$. (b)  The relative correction to the dimensionless radius of the innermost stable circular orbits $R_1/R_{0}$ with respect to $\frac{Q}{M}$ for various $\beta=0,\ 0.5,\ 1$, respectively.}
	\label{fig7}
\end{figure}
The graphic shows that  when the perturbation parameter $\epsilon >0$, the relative corrections to the radius of the innermost stable circular orbits are always negative, and the absolute value increases as  the charge-to-mass ratio increases.  On the other hand, with the increase of $\beta$, the relative corrections become smaller and smaller.

Now we consider the influence of $\beta$ on the radiative efficiency $\eta$ in the thin accretion disk model, which is defined by
\begin{equation}
\eta \equiv 1-{E}(R)\;.
\end{equation}
This quantity indicates the maximal fraction of energy being radiated when the test particle is accreted by a central black hole. The radiative efficiencies of the Schwarzschild black holes and extreme Kerr black holes are $0.057$ and $0.42$, respectively. From equation(\ref{isco1}), we know the that energy of ISCO is
\begin{equation}
E(R)\approx E(R_{0}+\epsilon R_{1},\epsilon)\approx E(R_{0})+\epsilon (R_{1}E_{1}(R_{0},0)+E_{2}(R_{0},0))\;,
\end{equation}
where $E_{1}$ and $E_{2}$ are the derivatives of $E(r,\epsilon)$ with respect to the first and second variables, respectively. Since  we have
\begin{equation}
\eta \approx 1-(E(R_{0})+\epsilon (R_{1}E_{1}(R_{0},0)+E_{2}(R_{0},0)))=1-E(R_{0})-\epsilon (R_{1}E_{1}(R_{0},0)+E_{2}(R_{0},0))\;,
\end{equation}
we can denote $1-E(R_{0})$ and $-(R_{1} E_{1}(R_{0},0)+E_{2}(R_{0},0))$ by $\eta_0$ and $\eta_1$, respectively.

\begin{figure}[htbp]
	
	\centering
	\subfigure[]{
		\includegraphics[width=8cm,height=6cm]{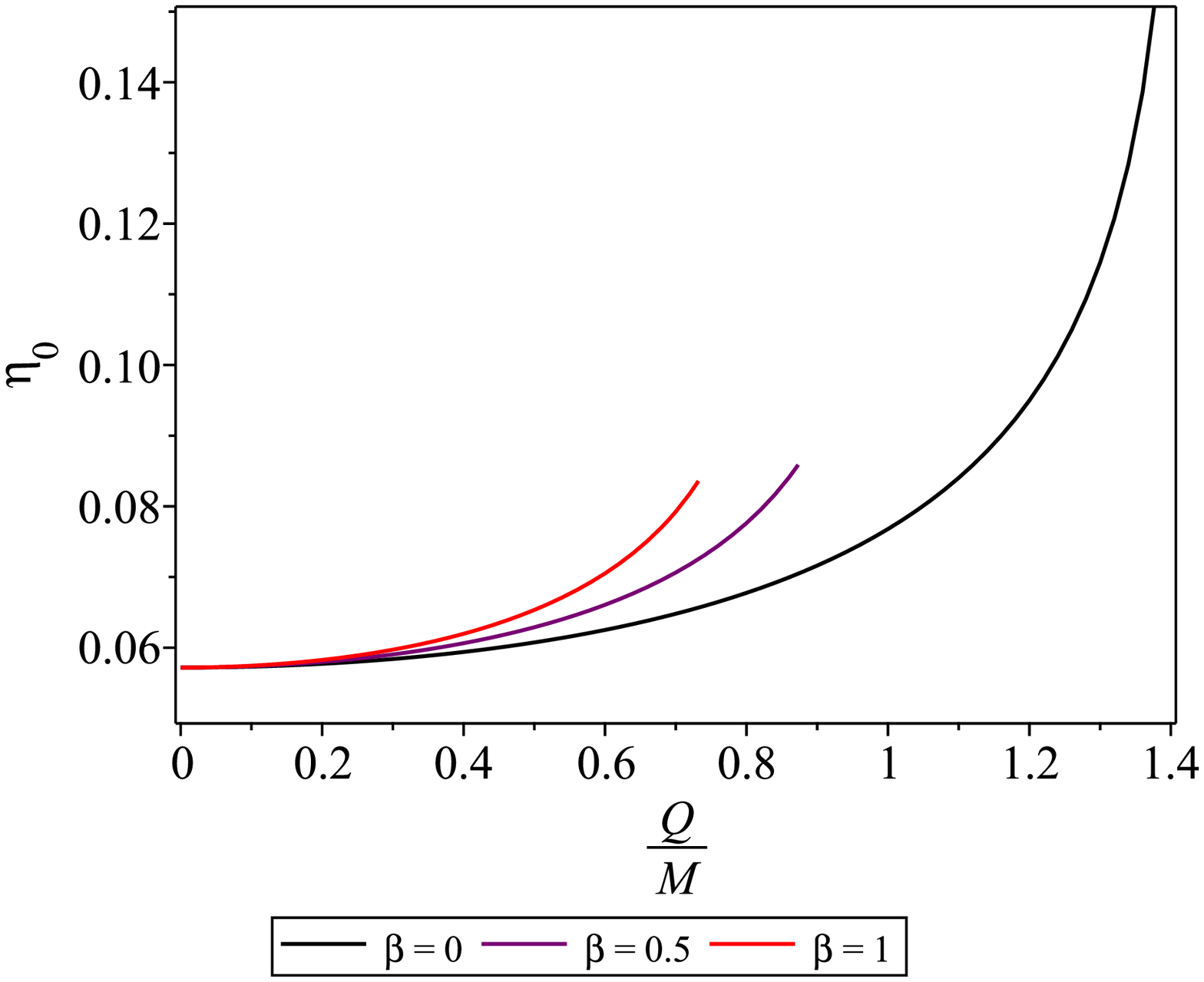}}
	\subfigure[]{
		\includegraphics[width=8cm,height=6cm]{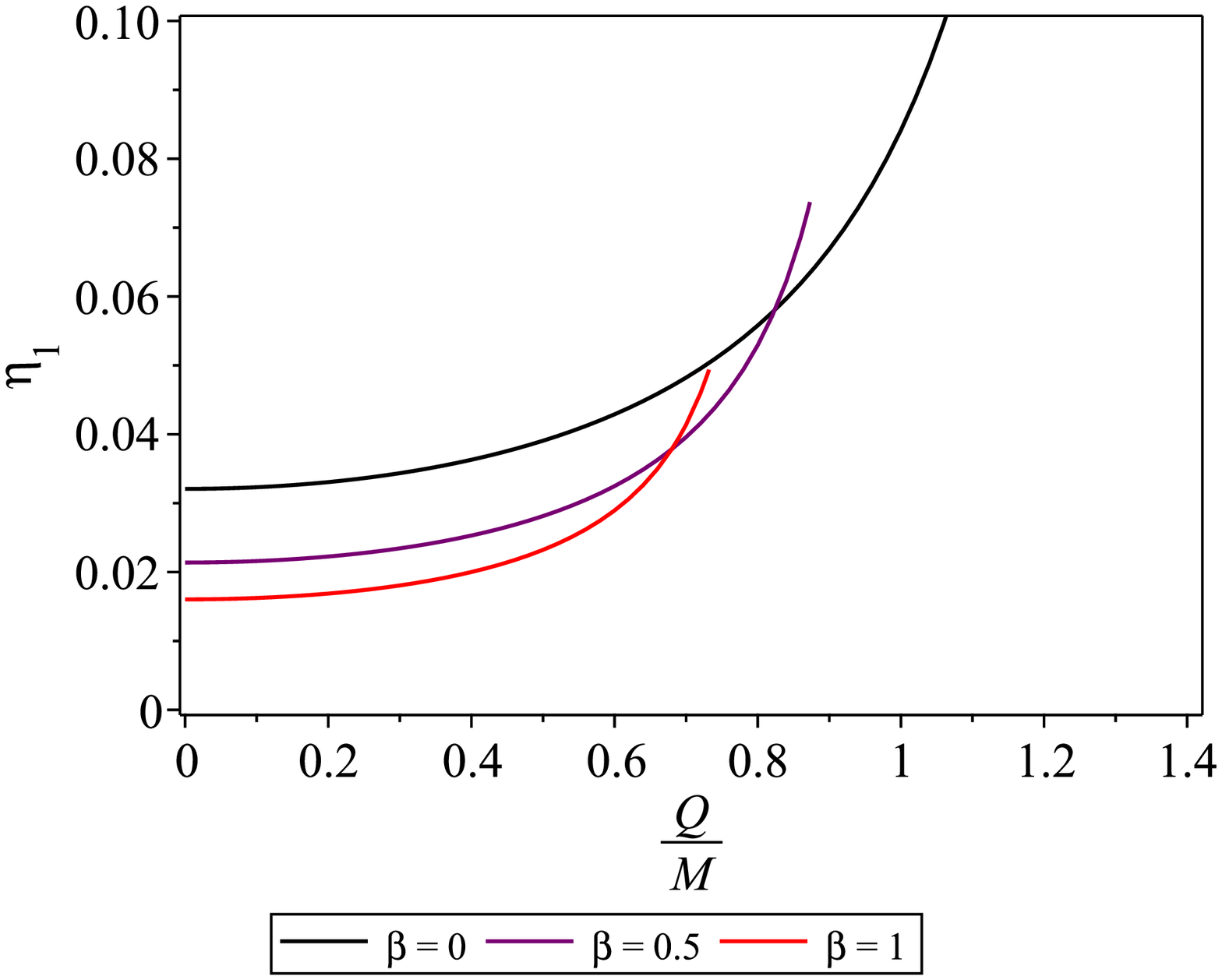}}\\
	\caption{(a) The radiation efficiency $\eta_0$ with respect to $\frac{Q}{M}$ for various $\beta$ in  static spacetime. (b) The first order corrections $\eta_1$ to radiation efficiency  with respect to $\frac{Q}{M}$ for various   $\beta$.}
	\label{fig8}
\end{figure}
The Fig.~\ref{fig8}(a) shows that the radiative efficiency $\eta_0$, starting from $1-\frac{2\sqrt{2}}{3} \approx 0.057$, increases with  $\frac{Q}{M}$ for a fixed value of $\beta$. When $\frac{Q}{M}$ is fixed, the radiative efficiency $\eta_0$ increases as $\beta$ increases.
Concerning the first order correction $\eta_1$, to the radiative efficiency, we see from the Fig.~\ref{fig8}(b) that the correction increases with regard to $\frac{Q}{M}$ when $\beta$ is fixed. For a fixed $\frac{Q}{M}$, the radiative efficiency $\eta_1$ decreases with the rise of $\beta$ in general, following the same trend as the correction to $R_{0}$.

Taking into account the zeroth order efficiency, $\eta_0$ and the first order efficiency, $\eta_1$, we obtain the total efficiency up to the first order in Fig.~\ref{fig9}.
\begin{figure}[htbp]
	\centering
	\includegraphics[width=8cm,height=6cm]{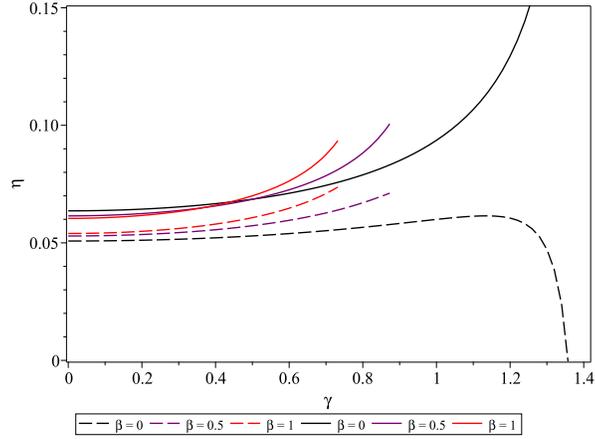}
	\caption{ The evolution of total efficiency $\eta$ up to first order with respect to $\frac{Q}{M}$ for different value of $\beta$. The solid and dashed lines correspond to $\epsilon=0.2$ and $\epsilon=-0.2$, respectively.}
	\label{fig9}
\end{figure}
It's found that for small values of $\beta$ and $\epsilon<0$, the efficiency will vanish for some value of $\frac{Q}{M}$. In this case, the black hole's capacity to capture particles becomes so weak that the accreted matter around it is greatly diluted. This leads to the failure of accretion disk to create radiation due to the insufficient amount of stresses and dynamical frictions.

 \section{Summary}
 In this paper, inspired by the categorization of EMS models, we investigate the Einstein-Maxwell-scalar theories $\mathcal{S}_{1}$ and $\mathcal{S}_{3}$  that admit both RN and dilaton solutions. The theories can also be obtained from EMD theories by an electromagnetic duality after omitting the axion term. They are classified as the scalarised-disconnected-type according to \cite{astefanesei2019einstein}. We conduct study on the slowly rotating black holes and summarize the key findings as follows.

\textbf{1.} The requirement of asymptotic flatness is not enough in order to determine the solution of the slowly rotating black hole. Therefore, to specify  the constant of integration for any value of $\beta$, the regular condition on the outermost event horizon must be imposed.

\textbf{2.} We constrain the range of $\beta$ by demanding that the black hole singularity is not naked and the coupling between the scalar field and the Maxwell field $K(\phi)$ is normal (not phantom). Then, we find the range of $\gamma$ cannot be $[0,2)$ for several values of $\beta$. What really intrigues us is the gyromagnetic ratio of the rotating black hole. We know the rotating charged black hole, i.e. the Kerr-Newman black hole, is noteworthy for having the
same gyromagnetic ratio of $2$ as the Dirac electron. Additionally, it is well established that the dilaton black hole always has a gyromagnetic ratio smaller than $2$, implying that the gyromagnetic ratio is suppressed in the presence of scalar hair. The model we investigate combines these two theories naturally and we are interested in its gyromagnetic ratio too. It's found that, when $b/a$ and $\beta$ are fixed, the gyromagnetic ratio decrease with the increase of $\gamma$. For fixed $b/a$ and $\gamma$, the gyromagnetic ratio also increases with increasing  $\beta$. Thus, by simultaneously raising $\beta$ and $\gamma$, the gyromagnetic ratio of the four-dimensional Kerr-Newman black hole can be obtained. Due to the presence of the second free parameter $\beta$, the suppressing of gyromagnetic ratio by the scalar hair may be overcome.

\textbf{3.} We have investigated the case of $\alpha=1$ as a specific example. The evolution of gyromagnetic ratio with respect to the charge-to-mass ratio is studied. We find that the gyromagnetic ratio increases with the increase of} $\beta$, and decreases with increasing $\frac{Q}{M}$. In view of this point, the gyromagnetic ratio may be restored by concurrently raising $\beta$ and $\frac{Q}{M}$. The dimensionless angular velocity of a locally non-rotating observer on the event horizon is also studied since it is related to the phenomenon of super-radiation. We find that the same value of the angular velocity as in the Kerr-Newman black hole can be obtained by increasing $\frac{Q}{M}$ and $\beta$.

\textbf{4.} Finally, the corrections to the period of circular orbits and the radius of the innermost stable circular orbits are  studied. It is found the relative correction decreases with increasing $\beta$ and increases with increasing $\frac{Q}{M}$. We also studied the radiative efficiency and the related correction in the thin accretion disk model. It's found that the black hole's capacity to capture particles becomes so weak that the accreted matter becomes very dilute provided that $\beta$ is very tiny and the perturbation parameter $\epsilon$ is negative. As a result, radiation cannot be created owing to the lack of sufficient stresses and dynamical frictions.

In all, the novel Einstein-Maxwell-scalar theory combines the well-known RN and dilaton black hole solutions. Additionally, the study on the slowly rotating black hole indicates it can mimic the properties of Kerr or Kerr-Newman black holes by adjusting the free parameter $\beta$. It is worth mentioning that a limitation of our work is that we don’t analyze the stability of the slowly rotating black
hole and this should be included in future work.

\section*{ACKNOWLEDGMENTS}
This work is partially supported by the Strategic Priority Research Program ``Multi-wavelength Gravitational Wave Universe'' of the
CAS, Grant No. XDB23040100 and the NSFC under grants 11633004, 11773031.

\section*{REFERENCES}

\bibliography{citationlist.bib}

\begin{thebibliography}{46}
\expandafter\ifx\csname natexlab\endcsname\relax\def\natexlab#1{#1}\fi
\expandafter\ifx\csname bibnamefont\endcsname\relax
  \def\bibnamefont#1{#1}\fi
\expandafter\ifx\csname bibfnamefont\endcsname\relax
  \def\bibfnamefont#1{#1}\fi
\expandafter\ifx\csname citenamefont\endcsname\relax
  \def\citenamefont#1{#1}\fi
\expandafter\ifx\csname url\endcsname\relax
  \def\url#1{\texttt{#1}}\fi
\expandafter\ifx\csname urlprefix\endcsname\relax\def\urlprefix{URL }\fi
\providecommand{\bibinfo}[2]{#2}
\providecommand{\eprint}[2][]{\url{#2}}

\bibitem[{\citenamefont{Rovelli}(1998)}]{rovelli1998strings}
\bibinfo{author}{\bibfnamefont{C.}~\bibnamefont{Rovelli}},
  \bibinfo{journal}{arXiv preprint gr-qc/9803024}  (\bibinfo{year}{1998}).

\bibitem[{\citenamefont{Cheung et~al.}(2008)\citenamefont{Cheung, Fitzpatrick,
  Kaplan, Senatore, and Creminelli}}]{cheung2008effective}
\bibinfo{author}{\bibfnamefont{C.}~\bibnamefont{Cheung}},
  \bibinfo{author}{\bibfnamefont{A.~L.} \bibnamefont{Fitzpatrick}},
  \bibinfo{author}{\bibfnamefont{J.}~\bibnamefont{Kaplan}},
  \bibinfo{author}{\bibfnamefont{L.}~\bibnamefont{Senatore}}, \bibnamefont{and}
  \bibinfo{author}{\bibfnamefont{P.}~\bibnamefont{Creminelli}},
  \bibinfo{journal}{Journal of High Energy Physics}
  \textbf{\bibinfo{volume}{2008}}, \bibinfo{pages}{014} (\bibinfo{year}{2008}).

\bibitem[{\citenamefont{Weinberg}(2008)}]{weinberg2008effective}
\bibinfo{author}{\bibfnamefont{S.}~\bibnamefont{Weinberg}},
  \bibinfo{journal}{Physical Review D} \textbf{\bibinfo{volume}{77}},
  \bibinfo{pages}{123541} (\bibinfo{year}{2008}).

\bibitem[{\citenamefont{Famaey and McGaugh}(2012)}]{famaey2012modified}
\bibinfo{author}{\bibfnamefont{B.}~\bibnamefont{Famaey}} \bibnamefont{and}
  \bibinfo{author}{\bibfnamefont{S.~S.} \bibnamefont{McGaugh}},
  \bibinfo{journal}{Living reviews in relativity}
  \textbf{\bibinfo{volume}{15}}, \bibinfo{pages}{10} (\bibinfo{year}{2012}).

\bibitem[{\citenamefont{Clifton et~al.}(2012)\citenamefont{Clifton, Ferreira,
  Padilla, and Skordis}}]{clifton2012modified}
\bibinfo{author}{\bibfnamefont{T.}~\bibnamefont{Clifton}},
  \bibinfo{author}{\bibfnamefont{P.~G.} \bibnamefont{Ferreira}},
  \bibinfo{author}{\bibfnamefont{A.}~\bibnamefont{Padilla}}, \bibnamefont{and}
  \bibinfo{author}{\bibfnamefont{C.}~\bibnamefont{Skordis}},
  \bibinfo{journal}{Physics reports} \textbf{\bibinfo{volume}{513}},
  \bibinfo{pages}{1} (\bibinfo{year}{2012}).

\bibitem[{\citenamefont{Joyce et~al.}(2016)\citenamefont{Joyce, Lombriser, and
  Schmidt}}]{joyce2016dark}
\bibinfo{author}{\bibfnamefont{A.}~\bibnamefont{Joyce}},
  \bibinfo{author}{\bibfnamefont{L.}~\bibnamefont{Lombriser}},
  \bibnamefont{and} \bibinfo{author}{\bibfnamefont{F.}~\bibnamefont{Schmidt}},
  \bibinfo{journal}{Annual Review of Nuclear and Particle Science}
  \textbf{\bibinfo{volume}{66}}, \bibinfo{pages}{95} (\bibinfo{year}{2016}).

\bibitem[{\citenamefont{Brans and Dicke}(1961)}]{brans1961mach}
\bibinfo{author}{\bibfnamefont{C.}~\bibnamefont{Brans}} \bibnamefont{and}
  \bibinfo{author}{\bibfnamefont{R.~H.} \bibnamefont{Dicke}},
  \bibinfo{journal}{Physical review} \textbf{\bibinfo{volume}{124}},
  \bibinfo{pages}{925} (\bibinfo{year}{1961}).

\bibitem[{\citenamefont{O'Hanlon}(1972)}]{o1972intermediate}
\bibinfo{author}{\bibfnamefont{J.}~\bibnamefont{O'Hanlon}},
  \bibinfo{journal}{Physical Review Letters} \textbf{\bibinfo{volume}{29}},
  \bibinfo{pages}{137} (\bibinfo{year}{1972}).

\bibitem[{\citenamefont{Acharya and Hogan}(1973)}]{acharya1973equivalence}
\bibinfo{author}{\bibfnamefont{R.}~\bibnamefont{Acharya}} \bibnamefont{and}
  \bibinfo{author}{\bibfnamefont{P.}~\bibnamefont{Hogan}},
  \bibinfo{journal}{Lettere al Nuovo Cimento (1971-1985)}
  \textbf{\bibinfo{volume}{6}}, \bibinfo{pages}{668} (\bibinfo{year}{1973}).

\bibitem[{\citenamefont{Fujii}(1971)}]{fujii1971dilaton}
\bibinfo{author}{\bibfnamefont{Y.}~\bibnamefont{Fujii}},
  \bibinfo{journal}{Nature Physical Science} \textbf{\bibinfo{volume}{234}},
  \bibinfo{pages}{5} (\bibinfo{year}{1971}).

\bibitem[{\citenamefont{Fujii and Maeda}(2003)}]{fujii2003scalar}
\bibinfo{author}{\bibfnamefont{Y.}~\bibnamefont{Fujii}} \bibnamefont{and}
  \bibinfo{author}{\bibfnamefont{K.-i.} \bibnamefont{Maeda}},
  \emph{\bibinfo{title}{The scalar-tensor theory of gravitation}}
  (\bibinfo{publisher}{Cambridge University Press}, \bibinfo{year}{2003}).

\bibitem[{\citenamefont{Sonner and Townsend}(2006)}]{sonner2006recurrent}
\bibinfo{author}{\bibfnamefont{J.}~\bibnamefont{Sonner}} \bibnamefont{and}
  \bibinfo{author}{\bibfnamefont{P.~K.} \bibnamefont{Townsend}},
  \bibinfo{journal}{Physical Review D} \textbf{\bibinfo{volume}{74}},
  \bibinfo{pages}{103508} (\bibinfo{year}{2006}).

\bibitem[{\citenamefont{Gibbons and Maeda}(1988)}]{gibbons1988black}
\bibinfo{author}{\bibfnamefont{G.~W.} \bibnamefont{Gibbons}} \bibnamefont{and}
  \bibinfo{author}{\bibfnamefont{K.-i.} \bibnamefont{Maeda}},
  \bibinfo{journal}{Nuclear Physics B} \textbf{\bibinfo{volume}{298}},
  \bibinfo{pages}{741} (\bibinfo{year}{1988}).

\bibitem[{\citenamefont{Garfinkle et~al.}(1991)\citenamefont{Garfinkle,
  Horowitz, and Strominger}}]{garfinkle1991charged}
\bibinfo{author}{\bibfnamefont{D.}~\bibnamefont{Garfinkle}},
  \bibinfo{author}{\bibfnamefont{G.~T.} \bibnamefont{Horowitz}},
  \bibnamefont{and}
  \bibinfo{author}{\bibfnamefont{A.}~\bibnamefont{Strominger}},
  \bibinfo{journal}{Physical Review D} \textbf{\bibinfo{volume}{43}},
  \bibinfo{pages}{3140} (\bibinfo{year}{1991}).

\bibitem[{\citenamefont{Cai and Zhang}(1996)}]{cai1996black}
\bibinfo{author}{\bibfnamefont{R.-G.} \bibnamefont{Cai}} \bibnamefont{and}
  \bibinfo{author}{\bibfnamefont{Y.-Z.} \bibnamefont{Zhang}},
  \bibinfo{journal}{Physical Review D} \textbf{\bibinfo{volume}{54}},
  \bibinfo{pages}{4891} (\bibinfo{year}{1996}).

\bibitem[{\citenamefont{Cai et~al.}(1998)\citenamefont{Cai, Ji, and
  Soh}}]{cai1998topological}
\bibinfo{author}{\bibfnamefont{R.-G.} \bibnamefont{Cai}},
  \bibinfo{author}{\bibfnamefont{J.-Y.} \bibnamefont{Ji}}, \bibnamefont{and}
  \bibinfo{author}{\bibfnamefont{K.-S.} \bibnamefont{Soh}},
  \bibinfo{journal}{Physical Review D} \textbf{\bibinfo{volume}{57}},
  \bibinfo{pages}{6547} (\bibinfo{year}{1998}).

\bibitem[{\citenamefont{Kunduri and Lucietti}(2005)}]{kunduri2005electrically}
\bibinfo{author}{\bibfnamefont{H.~K.} \bibnamefont{Kunduri}} \bibnamefont{and}
  \bibinfo{author}{\bibfnamefont{J.}~\bibnamefont{Lucietti}},
  \bibinfo{journal}{Physics Letters B} \textbf{\bibinfo{volume}{609}},
  \bibinfo{pages}{143} (\bibinfo{year}{2005}).

\bibitem[{\citenamefont{Kunz et~al.}(2006)\citenamefont{Kunz, Maison,
  Navarro-L{\'e}rida, and Viebahn}}]{kunz2006rotating}
\bibinfo{author}{\bibfnamefont{J.}~\bibnamefont{Kunz}},
  \bibinfo{author}{\bibfnamefont{D.}~\bibnamefont{Maison}},
  \bibinfo{author}{\bibfnamefont{F.}~\bibnamefont{Navarro-L{\'e}rida}},
  \bibnamefont{and} \bibinfo{author}{\bibfnamefont{J.}~\bibnamefont{Viebahn}},
  \bibinfo{journal}{Physics Letters B} \textbf{\bibinfo{volume}{639}},
  \bibinfo{pages}{95} (\bibinfo{year}{2006}).

\bibitem[{\citenamefont{Brihaye et~al.}(2007)\citenamefont{Brihaye, Radu, and
  Stelea}}]{brihaye2007black}
\bibinfo{author}{\bibfnamefont{Y.}~\bibnamefont{Brihaye}},
  \bibinfo{author}{\bibfnamefont{E.}~\bibnamefont{Radu}}, \bibnamefont{and}
  \bibinfo{author}{\bibfnamefont{C.}~\bibnamefont{Stelea}},
  \bibinfo{journal}{Classical and Quantum Gravity}
  \textbf{\bibinfo{volume}{24}}, \bibinfo{pages}{4839} (\bibinfo{year}{2007}).

\bibitem[{\citenamefont{Sen}(1992)}]{sen1992rotating}
\bibinfo{author}{\bibfnamefont{A.}~\bibnamefont{Sen}},
  \bibinfo{journal}{Physical Review Letters} \textbf{\bibinfo{volume}{69}},
  \bibinfo{pages}{1006} (\bibinfo{year}{1992}).

\bibitem[{\citenamefont{Hioki and Miyamoto}(2008)}]{hioki2008hidden}
\bibinfo{author}{\bibfnamefont{K.}~\bibnamefont{Hioki}} \bibnamefont{and}
  \bibinfo{author}{\bibfnamefont{U.}~\bibnamefont{Miyamoto}},
  \bibinfo{journal}{Physical Review D} \textbf{\bibinfo{volume}{78}},
  \bibinfo{pages}{044007} (\bibinfo{year}{2008}).

\bibitem[{\citenamefont{Pradhan}(2016)}]{pradhan2016thermodynamic}
\bibinfo{author}{\bibfnamefont{P.}~\bibnamefont{Pradhan}},
  \bibinfo{journal}{The European Physical Journal C}
  \textbf{\bibinfo{volume}{76}}, \bibinfo{pages}{131} (\bibinfo{year}{2016}).

\bibitem[{\citenamefont{Uniyal et~al.}(2017)\citenamefont{Uniyal, Nandan, and
  Purohit}}]{uniyal2017null}
\bibinfo{author}{\bibfnamefont{R.}~\bibnamefont{Uniyal}},
  \bibinfo{author}{\bibfnamefont{H.}~\bibnamefont{Nandan}}, \bibnamefont{and}
  \bibinfo{author}{\bibfnamefont{K.}~\bibnamefont{Purohit}},
  \bibinfo{journal}{Classical and Quantum Gravity}
  \textbf{\bibinfo{volume}{35}}, \bibinfo{pages}{025003}
  (\bibinfo{year}{2017}).

\bibitem[{\citenamefont{Delgado
  et~al.}(2016{\natexlab{a}})\citenamefont{Delgado, Herdeiro, and
  Radu}}]{delgado2016violations}
\bibinfo{author}{\bibfnamefont{J.~F.} \bibnamefont{Delgado}},
  \bibinfo{author}{\bibfnamefont{C.~A.} \bibnamefont{Herdeiro}},
  \bibnamefont{and} \bibinfo{author}{\bibfnamefont{E.}~\bibnamefont{Radu}},
  \bibinfo{journal}{Physical Review D} \textbf{\bibinfo{volume}{94}},
  \bibinfo{pages}{024006} (\bibinfo{year}{2016}{\natexlab{a}}).

\bibitem[{\citenamefont{Casadio et~al.}(1997)\citenamefont{Casadio, Harms,
  Leblanc, and Cox}}]{casadio1997new}
\bibinfo{author}{\bibfnamefont{R.}~\bibnamefont{Casadio}},
  \bibinfo{author}{\bibfnamefont{B.}~\bibnamefont{Harms}},
  \bibinfo{author}{\bibfnamefont{Y.}~\bibnamefont{Leblanc}}, \bibnamefont{and}
  \bibinfo{author}{\bibfnamefont{P.}~\bibnamefont{Cox}},
  \bibinfo{journal}{Physical Review D} \textbf{\bibinfo{volume}{55}},
  \bibinfo{pages}{814} (\bibinfo{year}{1997}).

\bibitem[{\citenamefont{Horne and Horowitz}(1992)}]{horne1992rotating}
\bibinfo{author}{\bibfnamefont{J.~H.} \bibnamefont{Horne}} \bibnamefont{and}
  \bibinfo{author}{\bibfnamefont{G.~T.} \bibnamefont{Horowitz}},
  \bibinfo{journal}{Physical Review D} \textbf{\bibinfo{volume}{46}},
  \bibinfo{pages}{1340} (\bibinfo{year}{1992}).

\bibitem[{\citenamefont{Shiraishi}(1992)}]{shiraishi1992spinning}
\bibinfo{author}{\bibfnamefont{K.}~\bibnamefont{Shiraishi}},
  \bibinfo{journal}{Physics Letters A} \textbf{\bibinfo{volume}{166}},
  \bibinfo{pages}{298} (\bibinfo{year}{1992}).

\bibitem[{\citenamefont{Sheykhi}(2007)}]{sheykhi2007thermodynamics}
\bibinfo{author}{\bibfnamefont{A.}~\bibnamefont{Sheykhi}},
  \bibinfo{journal}{Physical Review D} \textbf{\bibinfo{volume}{76}},
  \bibinfo{pages}{124025} (\bibinfo{year}{2007}).

\bibitem[{\citenamefont{Sheykhi et~al.}(2008)\citenamefont{Sheykhi,
  Allahverdizadeh, Bahrampour, and Rahnama}}]{sheykhi2008asymptotically}
\bibinfo{author}{\bibfnamefont{A.}~\bibnamefont{Sheykhi}},
  \bibinfo{author}{\bibfnamefont{M.}~\bibnamefont{Allahverdizadeh}},
  \bibinfo{author}{\bibfnamefont{Y.}~\bibnamefont{Bahrampour}},
  \bibnamefont{and} \bibinfo{author}{\bibfnamefont{M.}~\bibnamefont{Rahnama}},
  \bibinfo{journal}{Physics Letters B} \textbf{\bibinfo{volume}{666}},
  \bibinfo{pages}{82} (\bibinfo{year}{2008}).

\bibitem[{\citenamefont{Ghosh and SenGupta}(2007)}]{ghosh2007slowly}
\bibinfo{author}{\bibfnamefont{T.}~\bibnamefont{Ghosh}} \bibnamefont{and}
  \bibinfo{author}{\bibfnamefont{S.}~\bibnamefont{SenGupta}},
  \bibinfo{journal}{Physical Review D} \textbf{\bibinfo{volume}{76}},
  \bibinfo{pages}{087504} (\bibinfo{year}{2007}).

\bibitem[{\citenamefont{Ayzenberg and Yunes}(2014)}]{ayzenberg2014slowly}
\bibinfo{author}{\bibfnamefont{D.}~\bibnamefont{Ayzenberg}} \bibnamefont{and}
  \bibinfo{author}{\bibfnamefont{N.}~\bibnamefont{Yunes}},
  \bibinfo{journal}{Physical Review D} \textbf{\bibinfo{volume}{90}},
  \bibinfo{pages}{044066} (\bibinfo{year}{2014}).

\bibitem[{\citenamefont{Herdeiro et~al.}(2018)\citenamefont{Herdeiro, Radu,
  Sanchis-Gual, and Font}}]{herdeiro2018spontaneous}
\bibinfo{author}{\bibfnamefont{C.~A.} \bibnamefont{Herdeiro}},
  \bibinfo{author}{\bibfnamefont{E.}~\bibnamefont{Radu}},
  \bibinfo{author}{\bibfnamefont{N.}~\bibnamefont{Sanchis-Gual}},
  \bibnamefont{and} \bibinfo{author}{\bibfnamefont{J.~A.} \bibnamefont{Font}},
  \bibinfo{journal}{Physical review letters} \textbf{\bibinfo{volume}{121}},
  \bibinfo{pages}{101102} (\bibinfo{year}{2018}).

\bibitem[{\citenamefont{Fernandes et~al.}(2019)\citenamefont{Fernandes,
  Herdeiro, Pombo, Radu, and Sanchis-Gual}}]{fernandes2019spontaneous}
\bibinfo{author}{\bibfnamefont{P.~G.} \bibnamefont{Fernandes}},
  \bibinfo{author}{\bibfnamefont{C.~A.} \bibnamefont{Herdeiro}},
  \bibinfo{author}{\bibfnamefont{A.~M.} \bibnamefont{Pombo}},
  \bibinfo{author}{\bibfnamefont{E.}~\bibnamefont{Radu}}, \bibnamefont{and}
  \bibinfo{author}{\bibfnamefont{N.}~\bibnamefont{Sanchis-Gual}},
  \bibinfo{journal}{Classical and Quantum Gravity}
  \textbf{\bibinfo{volume}{36}}, \bibinfo{pages}{134002}
  (\bibinfo{year}{2019}).

\bibitem[{\citenamefont{Astefanesei et~al.}(2019)\citenamefont{Astefanesei,
  Herdeiro, Pombo, and Radu}}]{astefanesei2019einstein}
\bibinfo{author}{\bibfnamefont{D.}~\bibnamefont{Astefanesei}},
  \bibinfo{author}{\bibfnamefont{C.}~\bibnamefont{Herdeiro}},
  \bibinfo{author}{\bibfnamefont{A.}~\bibnamefont{Pombo}}, \bibnamefont{and}
  \bibinfo{author}{\bibfnamefont{E.}~\bibnamefont{Radu}},
  \bibinfo{journal}{Journal of High Energy Physics}
  \textbf{\bibinfo{volume}{2019}}, \bibinfo{pages}{1} (\bibinfo{year}{2019}).

\bibitem[{\citenamefont{Yu et~al.}(2021)\citenamefont{Yu, Qiu, and
  Gao}}]{yu2021constructing}
\bibinfo{author}{\bibfnamefont{S.}~\bibnamefont{Yu}},
  \bibinfo{author}{\bibfnamefont{J.}~\bibnamefont{Qiu}}, \bibnamefont{and}
  \bibinfo{author}{\bibfnamefont{C.}~\bibnamefont{Gao}},
  \bibinfo{journal}{Classical and Quantum Gravity}
  \textbf{\bibinfo{volume}{38}}, \bibinfo{pages}{105006}
  (\bibinfo{year}{2021}).

\bibitem[{\citenamefont{Turimov et~al.}(2020)\citenamefont{Turimov, Rayimbaev,
  Abdujabbarov, Ahmedov, and Stuchl{\'\i}k}}]{turimov2020test}
\bibinfo{author}{\bibfnamefont{B.}~\bibnamefont{Turimov}},
  \bibinfo{author}{\bibfnamefont{J.}~\bibnamefont{Rayimbaev}},
  \bibinfo{author}{\bibfnamefont{A.}~\bibnamefont{Abdujabbarov}},
  \bibinfo{author}{\bibfnamefont{B.}~\bibnamefont{Ahmedov}}, \bibnamefont{and}
  \bibinfo{author}{\bibfnamefont{Z.}~\bibnamefont{Stuchl{\'\i}k}},
  \bibinfo{journal}{Physical Review D} \textbf{\bibinfo{volume}{102}},
  \bibinfo{pages}{064052} (\bibinfo{year}{2020}).

\bibitem[{\citenamefont{Qiu and Gao}(2020)}]{qiu2020constructing}
\bibinfo{author}{\bibfnamefont{J.}~\bibnamefont{Qiu}} \bibnamefont{and}
  \bibinfo{author}{\bibfnamefont{C.}~\bibnamefont{Gao}},
  \bibinfo{journal}{Universe} \textbf{\bibinfo{volume}{6}},
  \bibinfo{pages}{148} (\bibinfo{year}{2020}).

\bibitem[{\citenamefont{Herdeiro and
  Oliveira}(2020)}]{herdeiro2020electromagnetic}
\bibinfo{author}{\bibfnamefont{C.~A.} \bibnamefont{Herdeiro}} \bibnamefont{and}
  \bibinfo{author}{\bibfnamefont{J.~M.} \bibnamefont{Oliveira}},
  \bibinfo{journal}{Journal of High Energy Physics}
  \textbf{\bibinfo{volume}{2020}}, \bibinfo{pages}{1} (\bibinfo{year}{2020}).

\bibitem[{\citenamefont{Barausse et~al.}(2016)\citenamefont{Barausse, Sotiriou,
  and Vega}}]{barausse2016slowly}
\bibinfo{author}{\bibfnamefont{E.}~\bibnamefont{Barausse}},
  \bibinfo{author}{\bibfnamefont{T.~P.} \bibnamefont{Sotiriou}},
  \bibnamefont{and} \bibinfo{author}{\bibfnamefont{I.}~\bibnamefont{Vega}},
  \bibinfo{journal}{Physical Review D} \textbf{\bibinfo{volume}{93}},
  \bibinfo{pages}{044044} (\bibinfo{year}{2016}).

\bibitem[{\citenamefont{Brown and York~Jr}(1993)}]{brown1993quasilocal}
\bibinfo{author}{\bibfnamefont{J.~D.} \bibnamefont{Brown}} \bibnamefont{and}
  \bibinfo{author}{\bibfnamefont{J.~W.} \bibnamefont{York~Jr}},
  \bibinfo{journal}{Physical Review D} \textbf{\bibinfo{volume}{47}},
  \bibinfo{pages}{1407} (\bibinfo{year}{1993}).

\bibitem[{\citenamefont{Gourgoulhon}(2012)}]{gourgoulhon20123+}
\bibinfo{author}{\bibfnamefont{E.}~\bibnamefont{Gourgoulhon}},
  \emph{\bibinfo{title}{3+ 1 formalism in general relativity: bases of
  numerical relativity}}, vol. \bibinfo{volume}{846}
  (\bibinfo{publisher}{Springer Science \& Business Media},
  \bibinfo{year}{2012}).

\bibitem[{\citenamefont{Delgado
  et~al.}(2016{\natexlab{b}})\citenamefont{Delgado, Herdeiro, Radu, and
  R{\'u}narsson}}]{delgado2016kerr}
\bibinfo{author}{\bibfnamefont{J.~F.} \bibnamefont{Delgado}},
  \bibinfo{author}{\bibfnamefont{C.~A.} \bibnamefont{Herdeiro}},
  \bibinfo{author}{\bibfnamefont{E.}~\bibnamefont{Radu}}, \bibnamefont{and}
  \bibinfo{author}{\bibfnamefont{H.}~\bibnamefont{R{\'u}narsson}},
  \bibinfo{journal}{Physics Letters B} \textbf{\bibinfo{volume}{761}},
  \bibinfo{pages}{234} (\bibinfo{year}{2016}{\natexlab{b}}).

\bibitem[{\citenamefont{Misner}(1972)}]{misner1972stability}
\bibinfo{author}{\bibfnamefont{C.}~\bibnamefont{Misner}},
  \bibinfo{journal}{Bulletin of the American Physical Society}
  \textbf{\bibinfo{volume}{17}}, \bibinfo{pages}{472} (\bibinfo{year}{1972}).

\bibitem[{\citenamefont{ZEL'DOVICH}(1972)}]{zel1972amplification}
\bibinfo{author}{\bibfnamefont{I.}~\bibnamefont{ZEL'DOVICH}},
  \bibinfo{journal}{Soviet Physics-JETP} \textbf{\bibinfo{volume}{35}},
  \bibinfo{pages}{1085} (\bibinfo{year}{1972}).

\bibitem[{\citenamefont{Starobinskii}(1973)}]{starobinskii1973amplification}
\bibinfo{author}{\bibfnamefont{A.}~\bibnamefont{Starobinskii}},
  \bibinfo{journal}{Zh. Eksp. Teor. Fiz} \textbf{\bibinfo{volume}{64}},
  \bibinfo{pages}{48} (\bibinfo{year}{1973}).

\bibitem[{\citenamefont{Shibata and Sasaki}(1998)}]{shibata1998innermost}
\bibinfo{author}{\bibfnamefont{M.}~\bibnamefont{Shibata}} \bibnamefont{and}
  \bibinfo{author}{\bibfnamefont{M.}~\bibnamefont{Sasaki}},
  \bibinfo{journal}{Physical Review D} \textbf{\bibinfo{volume}{58}},
  \bibinfo{pages}{104011} (\bibinfo{year}{1998}).

\end{thebibliography}

\end{document}